\documentclass[aps,prb,twocolumn,showpacs,preprintnumbers,amsmath,amssymb]{revtex4}

\usepackage{graphicx}
\usepackage{dcolumn}
\usepackage{bm}

\begin{document}

\title{Quantum Field Theory Solution for a Short-Range Interacting SO(3) Quantum
                        Spin-Glass \\}

\author{C.M.S. da Concei\c c\~ao$^1$ and E.C.Marino$^{2}$}
\affiliation{
 $^1$Departamento de F\'\i sica Te\'orica, Universidade do Estado do Rio de Janeiro, Rio de Janeiro, RJ 20550-013, Brazil}
\affiliation{$^2$ Instituto de F\'\i sica, Universidade Federal do
Rio de Janeiro, Caixa Postal 68528, Rio de Janeiro, RJ 21941-972,
Brazil}


\date{\today}

\begin{abstract}

 We study the quenched disordered magnetic system, which is obtained from the 2D SO(3) quantum
  Heisenberg model, on a square lattice, with nearest neighbors interaction, by taking a Gaussian
  random distribution of couplings centered in an antiferromagnetic coupling, $\bar J>0$ and with
  a width $\Delta J$. Using coherent spin states we can integrate over the random variables and map
   the system onto a field theory, which is a generalization of the SO(3) nonlinear sigma model with different flavors corresponding
   to the replicas, coupling parameter proportional to $\bar J$ and having a quartic spin interaction proportional
    to the disorder ( $\Delta J$). After deriving the CP$^1$ version of the system, we perform a calculation
    of the free energy density in the limit of zero replicas, which fully includes the quantum fluctuations
of the CP$^1$ fields $z_i$. We, thereby obtain the phase
    diagram of the system in terms of ($T, \bar J, \Delta J$). This presents an ordered antiferromagnetic (AF) phase,
     a paramagnetic (PM) phase and a spin-glass (SG) phase. A critical curve separating the PM and SG phases ends
      at a quantum critical point located between the AF and SG phases, at $T=0$. The Edwards-Anderson order
 parameter, as well as the magnetic susceptibilities are explicitly obtained in each of the three phases as a
 function of the three control parameters. The magnetic susceptibilities show a Curie-type behavior at high
 temperatures and exhibit a clear cusp, characteristic of the SG transition, at the transition line. The thermodynamic
 stability of the phases is investigated by a careful analysis of the Hessian matrix of the free energy. We show that
 all principal minors of the Hessian are positive in the limit of zero replicas, implying in particular that the SG
 phase is stable.
\end{abstract}

\maketitle

\section{Introduction}

\renewcommand{\theequation}{\arabic{section}.\arabic{equation}}
\setcounter{equation}{0}

A spin glass is a peculiar state, which is presented by certain disordered magnetic materials, when
the competition between opposite interactions produces what is known as frustration,
namely the incapability of the system to attain the lowest possible energy state that would correspond to each single type of
interaction. This situation is described by ascribing to the system a random distribution of coupling constants,
allowing interactions of opposite signs.
As a consequence of the competition between different types of order, some of the spin glass properties are
shared with paramagnetic states and some other with ordered, ferromagnetic or N\'eel states. Like the first ones, SG states possess
zero magnetization and similarly to ordered magnetic states, they present breakdown of ergodicity.
The time scale of disorder is typically much larger than the one associated to the dynamics and therefore
we must perform the quantum and thermal averages before averaging over the disorder,
the so called ``quenched'' thermodynamical description, which requires the use of the
replica method  \cite{by,mpv}.

A SG state exhibits clear theoretical and experimental signatures. The former has been introduced by Edwards and Anderson (EA),
who proposed a model for SG and devised an order parameter that captures one of the basic features of the glassy behavior,
namely the occurrence of
infinite time correlations for each spin  \cite{ea}. The same type of correlations exist in an ordered magnetic state such as
the N\'eel state or the ferromagnetic state, for instance. In the SG state, however, this happens without the associated existence
of infinite
spatial correlations among the spins and the consequent spontaneous nonzero magnetization (staggered in the case
of antiferromagnetic (AF) order).
Hence the SG state can be characterized as one presenting infinite time correlations (nonzero EA order parameter) but with
zero magnetization order parameter. From the experimental point of view, one of the most distinctive signatures
of the SG transition is a very sharp
cusp in the magnetic susceptibility as a function of the temperature, right at the transition \cite{by,mpv}.

An important landmark in the study of SG was the derivation of a
mean-field solution of a simplified version of the EA model,
obtained by Sherrington and Kirkpatrick  (SK)  \cite{sk}. They
considered a system with classical Ising spins with long-range
interactions, in which each spin would interact with any other spin
in the material, no matter how far apart they might be. This
assumption greatly facilitates, technically, the obtainment of the
mean-field solution. Soon after, however it was realized that the
solution of SK was unstable \cite{at}.  This fact has been generally
ascribed to the so-called replica symmetry, which was presented by
the SK solution. Indeed, later on, Parisi has found a stable replica
symmetry breaking solution \cite{par}.

The long-range interactions of the SK model, however, are likely to
be unphysical, to a large extent. Real materials should mostly be
 short-range interacting, quantum SO(3) Heisenberg spin systems. Despite the large amount of knowledge that we have today
about long-range interacting SG, surprisingly, very little is known
about the properties of short-range interacting quantum
spin-glasses, especially with SO(3) symmetry. Apart from some
numerical calculations \cite{renum}, very few analytical approaches
exist. An appealing short-range interacting disordered system, which
has been thoroughly investigated mostly by numerical methods is the
transverse field Ising spin glass \cite{ti}. Interesting related
results can be found in \cite{bm,persach} and \cite{rsy}, where a
Landau-Ginzburg, phenomenological approach has been developed for
the short-range SG problem.

We have proposed a model for a disordered SO(3) Heisenberg-like
quantum spin system with nearest neighbor interactions for which an
expression for the free energy density can be derived \cite{mm}. In
this work, we map the system onto a generalized CP$^1$ quantum field
theory in the continuum limit. This is a very convenient framework,
because nearest neighbor interactions become just derivatives. At
this point, however, we must be careful. Indeed, when taking the
continuum limit of a quantum system Berry phases will be generated,
which in general would not cancel when summed over the lattice
\cite{rsy,sy}. We may tackle this problem by introducing disorder as
a perturbation of an antiferromagnetic (AF) 2D Heisenberg model, for
which the sum of the quantum phases is known to cancel
\cite{xwef,ha1}.  Hence, we consider a Gaussian random distribution
of couplings centered in an AF coupling $\bar J
> 0$ and with variance $\Delta J$, such that $\Delta J \ll \bar J$.
The situation is completely different from the original EA model,
where $\bar J =0$, and consequently the disorder cannot be taken as
a perturbation of a Heisenberg system \cite{ea}.

Using the CP$^1$ description, we obtain a solution, which presents
replica symmetry. Out of this, we extract the $T\times \bar J$ phase
diagram of the system. This exhibits a N\'eel phase at $T=0$ and
$\rho_s > \rho_0$, where $\rho_s = S^2 \bar J$ is the spin stiffness
($S$ is the spin quantum number) and $\rho_0$ is a quantum critical
point. This is displaced by disorder to the right of  its original
value $\rho_0(0) =\frac{\Lambda}{2\pi}$ ($\Lambda=\frac{1}{a}$; $a$:
lattice spacing) in the pure system \cite{chn,em}. It also contains
a spin glass (SG) phase for temperatures below a certain critical
line $T < Tc$ and $\rho_s < \rho_0$, characterized by a nonzero EA
order parameter and zero staggered magnetization. For $T>0$ ;
$\rho_s > \rho_0$ and also for $T > Tc$ ; $\rho_s < \rho_0$, we find
a paramagnetic (PM) phase. The behavior of the magnetic
susceptibility, of the spin-gap and of the EA order parameter are
analyzed in detail in each phase as well as on the transition line.
The former presents a nice cusp at the transition, in agreement with
the typical experimental behavior of spin glasses \cite{by}. We also
show how the phase diagram is modified by varying the amount of
disorder

We make a thorough investigation about the thermodynamic stability
of our solution. This is done through a careful analysis of the
Hessian matrix of the average free-energy density. We show in detail
that all principal minors of the Hessian matrix are strictly
positive in the physical limit when the number of replicas reduces
to zero. This is a necessary and sufficient condition for the
mean-field solution to be a local minimum, hence it guarantees the
solution is stable. Furthermore, we can follow the phase transition
directly in the Hessian, as the corresponding principal minor
determinants cease to be positive if we use the wrong solution for a
given phase.

There is an appealing physical motivation for our SG model, in
connection to the high-Tc cuprates. Indeed, these materials, when
undoped, are 2D Heisenberg antiferromagnets, which upon doping,
develop a SG phase before becoming superconductors. Our model, also
being a 2D Heisenberg antiferromagnet for $\Delta =0$, describes
precisely the AF-SG transition as we increase the disorder and is
therefore potentially very useful for studying the magnetic
fluctuations of such materials.



\section{The Model and the Continuum Limit}

\setcounter{equation}{0}

\subsection{The Model}

The model consists of
an SO(3) quantum Heisenberg-like hamiltonian,
containing only nearest neighbor interactions of the spin operators $\mathbf{\widehat{S}}_{i}$, on the sites of a $2D$ square lattice
of spacing $a$,
\begin{equation}
\widehat{\mathcal{H}}=\sum_{\langle ij\rangle} J_{ij}
\mathbf{\widehat{S}}_{i}\cdot\mathbf{\widehat{S}}_{j},\label{Hpuro}
\end{equation}
The couplings $J_{ij}$ are random and associated with a Gaussian probability distribution
with variance $\Delta J$ and
centered in $\bar{J}>0$, such that $\Delta J\ll\bar J$, namely,
\begin{equation}
P[J_{ij}]=\frac{1}{\sqrt{2\pi(\Delta
J)^{2}}}\exp\left[-\frac{(J_{ij}-\bar{J})^{2}}{2(\Delta
J)^{2}}\right]\label{distgauss},
\end{equation}
We consider the quenched situation, in which, according to the replica method \cite{by,ea} the average free-energy is given by
\begin{equation}
\overline{F}=-k_{B}T\lim_{n\longrightarrow
0}\frac{1}{n}\left([Z^{n}]_{av}-1\right),
\end{equation}
where $Z^{n}$ is the replicated partition function for a given configuration of couplings $J_{ij}$,

\begin{equation}
Z^{n}\{J_{ij}\}=\mathop{\mathrm{Tr}}\limits^{
}_{\{\widehat{S}_{i}^{\alpha}\}}\exp\left[-\beta\sum_{\alpha=1}^{n}\sum_{\langle ij\rangle} J_{ij}
\mathbf{\widehat{S}}^\alpha_{i}\cdot\mathbf{\widehat{S}}^\alpha_{j}\right],\label{Znpuro}
\end{equation}
and
\begin{equation}
[Z^{n}]_{av}=\int\left(\prod_{(ij)}
dJ_{ij}P[J_{ij}]\right)Z^{n}\{J_{ij}\}.\label{zzinho}
\end{equation}

is the average thereof with the Gaussian distribution.

We now make use of the coherent spin states
$
|\mathbf{\Omega}(\tau)\rangle $, such that
\begin{eqnarray}
|\mathbf{\Omega}(\tau)\rangle\equiv\bigotimes_{i}\bigotimes_{\alpha=1}^{n}
|\mathbf{\Omega}_{i}^{\alpha}(\tau)\rangle.\nonumber
\end{eqnarray}
with
\begin{equation}
\langle\mathbf{\Omega}_{i}^{\alpha}(\tau)|\mathbf{\widehat{S}}^\alpha_{i}
|\mathbf{\Omega}_{i}^{\alpha}(\tau)\rangle = S \mathbf{\Omega}_{i}^{\alpha}(\tau),
\end{equation}
where $i$: lattice sites, $\alpha$: replicas, $\tau$: euclidian time, $S$: spin quantum number,
 \cite{ha1}, and $\mathbf{\Omega}_{i}^{\alpha}(\tau)$ is a classic vector of unit magnitude. With the help of these
coherent states
we may express $Z^{n}$  as a functional integral over
$\mathbf{\Omega}_{i}^{\alpha}(\tau)$, namely,
\begin{equation}
Z^{n}\{J_{ij}\}=\int\mathcal{D}\mathbf{\Omega} \exp\left\{- \int_0^\beta d\tau \sum_{\alpha=1}^{n} L^\alpha(\tau)\right\},
\end{equation}
where
$L^\alpha(\tau)$
\begin{equation}
L^\alpha(\tau)=L_B^\alpha
-S^{2}\sum_{\langle
ij\rangle}J_{ij}\mathbf{\Omega}_{i}^{\alpha}(\tau)\cdot\mathbf{\Omega}_{j}^{\alpha}(\tau)\label{hcoerente}.
\end{equation}
and
\begin{equation}
L_B^\alpha=\sum_{i} \langle\mathbf{\Omega}_{i}^{\alpha}(\tau)|\frac{d}{d\tau}
|\mathbf{\Omega}_{i}^{\alpha}(\tau)\rangle,
\end{equation}
is the Berry phase term.

The average over the disordered couplings $J_{ij}$
can then be performed, yielding
\begin{equation}
[Z^{n}]_{av}=\int\mathcal{D}\mathbf{\Omega} \exp\left\{- S_{\bar J,\Delta}\right\};
\end{equation}
where $S_{\bar J,\Delta}$ is given by
$$
S_{\bar J,\Delta}= \int_0^\beta d\tau\sum_\alpha \left[ L_B^\alpha
-S^{2}\bar J \sum_{\langle
ij\rangle}\mathbf{\Omega}_{i}^{\alpha}(\tau)\cdot\mathbf{\Omega}_{j}^{\alpha}(\tau)\right]
$$
\begin{equation}
+\frac{S^4 (\Delta J)^2}{2}\sum_{\langle ij\rangle}\int_0^\beta
d\tau d\tau' \Omega_{ia}^{\alpha}(\tau) \Omega_{ib}^{\beta}(\tau')
\Omega_{ja}^{\alpha}(\tau) \Omega_{jb}^{\beta}(\tau')
\label{hcoerente}
\end{equation}
where summations in $(\alpha \beta)$, as well as over the SO(3) components of $\mathbf{\Omega}$, $(ab)$ are understood.

Using the connectivity matrix, defined as:  ($K_{ij}=1$ if $(ij)$ are nearest neighbors and $K_{ij}=0$ otherwise), we may write the
nearest neighbor sums in the last term of (\ref{hcoerente}) as overall sums. This allows us to perform
a Hubbard-Stratonovitch transformation  that replaces the quartic term of (\ref{hcoerente}) by
$$
S_{\Delta} =
S^{4}(\Delta J)^{2}\sum_{\alpha,\beta=1}^{n} \int_0^\beta d\tau d\tau'\sum_{\langle
ij\rangle}\left [ \frac{1}{2} Q_{i,ab}^{\alpha\beta}(\tau,\tau')Q_{j,ab}^{\alpha\beta}(\tau,\tau')\right.
$$
\begin{equation}
\left.
-\Omega_{i,a}^{\alpha}(\tau)Q_{j,ab}^{\alpha\beta}(\tau,\tau')\Omega_{i,b}^{\beta}(\tau')\right].
\label{hamiltonianoefetivo1}
\end{equation}
This is no longer a disordered system. The  disorder, which was originally present manifests now through the interaction term,
proportional to $(\Delta J)^{2}$. In the absence of disorder, we would have $\Delta J \rightarrow 0$ and $[Z^{n}]_{av} $ would reduce
to the usual coherent spin representation of the AF Heisenberg model, with a coupling $\bar J > 0$ \cite{sach,chn,em}.


\subsection{Continuum Limit}

Since we are only considering the weakly disordered case $(\Delta J \ll \bar J)$ our model, described by the
effective action in (\ref{hcoerente}),
is a perturbation of the AF 2D quantum Heisenberg model. This means we can decompose the classical spin
$\Omega_{i,a}^{\alpha}(\tau)$ into antiferromagnetic and ferromagnetic fluctuations as in that model \cite{sach}. Using this, then
it follows that the sum of the quantum Berry phases,
$ L^B_{i,\alpha}$, over all the lattice sites vanishes, as in the pure system \cite{ha1,xwef}.

We can therefore take the continuum limit in the usual way as in the pure AF 2D quantum Heisenberg model \cite{ha1,ha2,sach}
obtaining an SO(3) generalized relativistic nonlinear sigma model (NLSM) describing the field $\mathbf{n}^\alpha=(\sigma^\alpha, \vec \pi^\alpha)$, which is
the continuum limit of the (staggered) spin $\mathbf{\Omega}^\alpha$:
\begin{eqnarray}
\mathcal{L}&=&\frac{1}{2}|\nabla\mathbf{n}^{\alpha}|^{2}+
\frac{1}{2c^{2}}|\partial_{\tau}\mathbf{n}^{\alpha}|^{2}+i\lambda_{\alpha}(|\mathbf{n}^{\alpha}
|^{2}-\rho_{s}) \nonumber  \\
&&+\frac{D}{2}
\left[Q_{ab}^{\alpha\beta}(\mathbf{r},\tau,\tau^{\prime})Q_{ab}^{\alpha\beta}(\mathbf{r},\tau,\tau^{\prime})\right.\label{acaoparadecompor}  \\
&&-\left.\frac{2}{\rho_{s}}n_{a}^{\alpha}(\mathbf{r},\tau)Q_{ab}^{\alpha\beta}(\mathbf{r},\tau,\tau^{\prime})n_{b}^{\beta}(\mathbf{r},\tau^{\prime})\right].\nonumber
\end{eqnarray}
where $D=S^4(\Delta J)^2/a^2$ ($a$: lattice parameter) and $\rho_s=S^2 \bar J$.
The constraint
$\mathbf{n}^\alpha\cdot\mathbf{n}^\alpha= \rho_s $ (no sum in $\alpha$), as usual is implemented by the Lagrange multiplier field $\lambda_\alpha$.

This generalized NLSM contains a $(\Delta J)^2$-proportional
 trilinear interaction of $n_a^\alpha$ with the
 Hubbard-Stratonovitch field $  Q_{ab}^{\alpha\beta}(\tau,\tau')$, which corresponds to the
 $\Delta J$-proportional part of (\ref{hcoerente}), namely $S_\Delta$.

Notice that a null
 value for $\bar J$, as we have in the EA model \cite{ea} would make the perturbation around a NLSM meaningless. A negative value, on
  the other hand, would correspond
 to the ferromagnetic Heisenberg model, which
 after taking the continuum limit, is associated to the
 non-relativistic NLSM. Here perturbation would be possible, however, the Berry's phases would no longer cancel. We emphasize, therefore, the enormous difference that exists, both from the physical
 and mathematical points of view, in considering $\bar J$ as positive, negative or null in the Gaussian distribution of the EA model.

The field $\mathbf{n}^\alpha=(\sigma^\alpha, \vec \pi^\alpha)$ is
the continuum limit of the (staggered) spin $\mathbf{\Omega}^\alpha$
and satisfies the constraint
$\mathbf{n}^\alpha\cdot\mathbf{n}^\alpha= \rho_s $, which is
implemented by integration on $\lambda^\alpha$.

Decomposing $Q^{\alpha\beta}$ into replica diagonal and off-diagonal
parts,
 \begin{equation}
 Q_{ab}^{\alpha\beta}(\vec r;\tau,\tau')\equiv \delta_{ab}[\delta^{\alpha\beta}\chi(\vec r;\tau,\tau')+
q^{\alpha\beta}(\vec r;\tau,\tau')]
\end{equation}
where $q^{\alpha\beta}=0$ for $\alpha = \beta$, we get

\begin{eqnarray}
\mathcal{L}_{\bar
J,\Delta}&=&\frac{1}{2}|\nabla\mathbf{n}^{\alpha}|^{2}+
\frac{1}{2c^{2}}|\partial_{\tau}\mathbf{n}^{\alpha}|^{2}+i\lambda_{\alpha}(|\mathbf{n}^{\alpha}
|^{2}-\rho_{s}) \nonumber  \\
&&+\frac{3D}{2} \int d\tau'
\left[n\chi^{2}(\tau,\tau^{\prime})+ q^{\alpha\beta}(\tau,\tau^{\prime})q^{\alpha\beta}(\tau,\tau^{\prime})\right.\nonumber  \\
&&-\left.\frac{D}{\rho_{s}}\mathbf{n}^{\alpha}(\tau)\chi(\tau,\tau^{\prime})
\mathbf{n}^{\alpha}(\tau^{\prime})\right.\nonumber  \\
&&-\left.\frac{D}{\rho_{s}}\mathbf{n}^{\alpha}(\tau)q^{\alpha\beta}(\tau,\tau^{\prime})
\mathbf{n}^{\beta}(\tau^{\prime})\right].\label{acaoparadecompor1}
\end{eqnarray}

This will be our starting point for the CP$^1$ formulation. In a
previous work \cite{mm}, we took a different path. From
(\ref{acaoparadecompor1}), we integrated over the $\vec \pi$-field
and thereby obtained an effective action for the remaining fields.

\section{ The CP$^1$ Formulation}
\setcounter{equation}{0}

\subsection{ CP$^1$ Lagranagian}

We now introduce the CP$^1$ field in the usual way, namely,
\begin{equation}
\mathbf{n}^{\alpha}(\tau) =
\frac{1}{\sqrt{\rho_s}}\left[z_i^{*\alpha}(\tau)\mathbf{\sigma}_{ij}
z_j^{\alpha}(\tau) \right]\label{nz}
\end{equation}
where the $z_i^\alpha$ field satisfies the constraint
\begin{equation}
|z_1^{\alpha}|^2 + |z_2^{\alpha}|^2 = \rho_s.\label{vinc}
\end{equation}
Using the above two equations, we get the correspondence
\begin{equation}
\frac{1}{2}|\nabla\mathbf{n}^{\alpha}|^{2}+
\frac{1}{2c^{2}}|\partial_{\tau}\mathbf{n}^{\alpha}|^{2}
\Leftrightarrow 2 \sum_{i=1}^2|D_\mu z_i^{\alpha}|^2 \label{01},
\end{equation}
where $D_\mu = \partial_\mu + i A_\mu$. The above correspondence
involves the functional integration over the auxiliary vector field
 $A_\mu$.

 Using (\ref{nz}), (\ref{vinc}) and (\ref{01}) in
(\ref{acaoparadecompor1}), we may express the average replicated
partition function in terms of the CP$^1$ fields as
\begin{equation}
[Z^{n}]_{av}=\int\mathcal{D}z  \mathcal{D}z^* \mathcal{D} A_\mu
\mathcal{D}\chi \mathcal{D}q \mathcal{D}\lambda e^{- S}\label{Z1},
\end{equation}
where $ S\left[z_i^\alpha, z_i^{\alpha *}, A_\mu, \lambda,
\chi(\tau,\tau'), q^{\alpha\beta}(\tau,\tau') \right]$ is the action
corresponding to the lagrangian density
$$
\mathcal{L}_{\bar J,\Delta,\mathbf{CP^1}}=2 |D_\mu z_i^{\alpha}|^2
+i\lambda_{\alpha}(|z_i^{\alpha}|^2-\rho_{s})
$$
$$
+\frac{3D}{2} \int d\tau' \left[n\chi^{2}(\tau,\tau^{\prime})+
q^{\alpha\beta}(\tau,\tau^{\prime})q^{\alpha\beta}(\tau,\tau^{\prime})\right]
$$
$$
+\frac{2D}{\rho^2_{s}} \int d\tau' \left
\{[\chi(\tau,\tau^{\prime})][|z_i^{*\alpha}(\tau)|^2
|z_j^{\alpha}(\tau^{\prime})|^2] \right.
$$
\begin{equation}
-\left.   [z_i^{*\alpha}
z_j^{\alpha}(\tau)][\chi(\tau,\tau^{\prime})\delta^{\alpha\beta}+
q^{\alpha\beta}(\tau,\tau^{\prime})] [z_i^{\beta}
z_j^{*\beta}(\tau^{\prime})] \right \},\label{acaoparadecompor11}
\end{equation}
where summation in $i,j,\alpha,\beta$ is understood.

\subsection{The Quantum Average Free Energy}

In order to evaluate $[Z^{n}]_{av}$ in (\ref{Z1}), we use the
stationary phase approximation. For this, we expand
$S\left[z_i^\alpha, z_i^{\alpha *}, A_\mu, \lambda,
\chi(\tau,\tau'), q^{\alpha\beta}(\tau,\tau') \right]$ around the
fields in the stationary point, in such a way that the quadratic
fluctuations about the $z_i^\alpha$ fields are taken into account,
namely,

$$
S\left[z_i^\alpha, z_i^{\alpha *}, A_\mu, \lambda_\alpha,  \chi,
q^{\alpha\beta} \right] =
 S\left[z_{i,\mathrm{s}}^\alpha, z_{i,\mathrm{s}}^{\alpha *}, A_\mu^\mathrm{s},  m^2,  \chi_{\mathrm{s}}, q_{\mathrm{s}}^{\alpha\beta} \right]
 $$
 \begin{equation}
  + \frac{1}{2}  \int d\tau d\tau' \eta_i^{\alpha *}(\tau) \mathbb{M}^{\alpha\beta}_{ij}(\tau,\tau') \eta_j^{\beta } (\tau') \label{Z2},
\end{equation}
where $\eta_i^{\alpha } = z_i^{\alpha } - z_{i,\mathrm{s}}^{\alpha
}$ and $\mathbb{M}$ is the matrix
\begin{eqnarray}
 \mathbb{M} =
\left(
\begin{array}{c}
\frac{\delta^2 S}{\delta z_i^\alpha(\tau) \delta z_j^{*\beta}(\tau^\prime)} \ \ \ \ \   \frac{\delta^2 S}{\delta z_i^{\alpha}(\tau) \delta z_j^{\beta}(\tau^\prime)} \\
\frac{\delta^2 S}{\delta z_i^{*\alpha}(\tau) \delta
z_j^{*\beta}(\tau^\prime)} \ \ \ \ \   \frac{\delta^2 S}{\delta
z_i^{*\alpha}(\tau) \delta z_j^{\beta}(\tau^\prime)}
\end{array}\right)\;\label{mat},
 \label{m}
 \end{eqnarray}
 with elements taken at the stationary fields.

These stationary fields are such that  $\lambda
^\alpha(\mathbf{r},\tau) \rightarrow \lambda_{\mathrm{s}}$ and $m^2
= 2i\lambda_{\mathrm{s}}$, which turns out to be the spin gap. Also
$\chi(\mathbf{r},\tau,\tau')\rightarrow
\chi_{\mathrm{s}}(\tau-\tau')$ and
$q^{\alpha\beta}(\mathbf{r},\tau,\tau')\rightarrow
q_{\mathrm{s}}^{\alpha\beta}(\tau-\tau')$. The staggered
magnetization $\sigma_{\mathrm{s}}^\alpha$ is given in terms of the
CP$^1$ fields as
\begin{equation}
\sigma^2_{\mathrm{s}} = \frac{1}{n} \sum_{\alpha=1}^n
\left[|z_{1,\mathrm{s}}^\alpha |^2+|z_{2,\mathrm{s}}^\alpha |^2
\right] \equiv \frac{1}{n} \sum_{\alpha=1}^n \sigma^2_\alpha .
\end{equation}
Finally, the stationary value of the gauge field is
$A^{\mathrm{s}}_\mu =0$.

Inserting (\ref{Z2}) in (\ref{Z1}) we obtain, after integrating over
the $z$-fields,
\begin{equation}
[Z^{n}]_{av}= e^{-
n{S}_{\mathrm{eff}}\left[\sigma_{\mathrm{s}}^\alpha, m^2,
A_\mu^\mathrm{s}=0, q_{\mathrm{s}}^{\alpha\beta}(\tau-\tau'),
\chi_{\mathrm{s}}(\tau-\tau')\right]},\label{funcaoparticaosigma}
\end{equation}
where
$$
 S_{\mathrm{eff}}\left[\sigma_{\mathrm{s}}^\alpha, m^2, A_\mu^\mathrm{s}=0, q_{\mathrm{s}}^{\alpha\beta}, \chi_{\mathrm{s}}\right]=  \beta V  \left \{  \frac{m^2}{2}\left[\sigma^2_{\mathrm{s}}-{\rho}_{s}\right]
    \right\} +
$$
$$
 + V\int_0^\beta d\tau d\tau^\prime \left \{ \frac{3D}{2}
\left[\chi^{2}_{\mathrm{s}}(\tau-\tau^{\prime})+ \frac{1}{n}
q^{\alpha\beta}_{\mathrm{s}}(\tau-\tau^{\prime})q^{\alpha\beta}_{\mathrm{s}}(\tau-\tau^{\prime})\right.\right.
$$
\begin{equation}
\left.-\frac{D}{n
\rho_s}\left[\chi_{\mathrm{s}}(\tau-\tau^{\prime})\delta^{\alpha\beta}+q^{\alpha\beta}_{\mathrm{s}}(\tau-\tau^{\prime})\right]
\sigma_{\alpha}\sigma_{\beta} \right \} -\frac{1}{n } \ln
\mathop{\mathrm{Det}} \mathbb{M}, \label{seff1}
 \end{equation}

Notice that the third term in (\ref{acaoparadecompor11}) is
proportional to $n^2 \sigma_{\mathrm{s}}^4$ and, therefore, does not
contribute to (\ref{seff1}) in the limit $n\rightarrow 0$.

By taking the limit $n\rightarrow 0$ in (\ref{funcaoparticaosigma})
we immediately realize that the average free-energy is given by
\begin{equation}
\bar F= \frac{1}{\beta}
S_{\mathrm{eff}}\left[\sigma_{\mathrm{s}}^\alpha, m^2,
A_\mu^\mathrm{s}=0, q_{\mathrm{s}}^{\alpha\beta}(\tau-\tau'),
\chi_{\mathrm{s}}(\tau-\tau')\right].
\end{equation}

In Appendix A we consider the determinant appearing in the last term
of (\ref{seff1}) This contains the quantum corrections coming from
the $z_i^\alpha$ fields. This determinant runs over the $i$
components of these fields, over the replicas and over the field
configurations. We are able to exactly calculate the first two
determinants by replacing the $q^{\alpha\beta}$ variables in the
last term of (\ref{acaoparadecompor11}) by their average,
\begin{equation}
\bar q =\lim_{n\rightarrow 0} \frac{1}{n(n-1)} \sum_{\alpha\beta}
q^{\alpha\beta}\label{qbar}
\end{equation}

The determinant over the field configurations is most conveniently
expressed in the space of Matsubara frequencies $\omega_r=2\pi rT, r
\in \mathbb{Z}$. For this purpose, we perform the Fourier
transformation of the  $\chi$'s and $q$'s. Then, using the
expression obtained for the determinant of the quantum fluctuations
(\ref{det111}), we get the average free-energy density as the
functional (we henceforth neglect the ``${\mathrm{s}}$'' subscript)
$$
 \bar f \left[\sigma^\alpha, m^2, q^{\alpha\beta}(\omega_r), \chi(\omega_r)\right]
= \frac{m^2}{2}\left[\sigma^2-{\rho}_{s}\right]
$$
$$
 -\frac{D}{n \rho_s}\left[\chi(\omega_0)\delta^{\alpha\beta}+q^{\alpha\beta}(\omega_0)\right]
\sigma^{\alpha}\sigma^{\beta}
$$
$$
+3DT\sum_{\omega_r}\left[ \chi(-\omega_r)\chi(\omega_r)
 + \frac{1}{n}q^{\alpha\beta}(-\omega_r)q^{\alpha\beta}(\omega_r)\right ]\label{fe2}
$$
\begin{equation}
+T\sum_{\omega_{r}}\int\frac{d^{2}k}{(2\pi)^{2}}\left[\ln\left(k^{2}+M_r\right)
-\frac{A\bar q(\omega_{m})}{k^{2}+M_r }\right] \label{f}
\end{equation}
where $A=\frac{2D}{\rho_s}$ and
\begin{equation}
M_{r}\equiv M(\omega_{r})=\omega_{r}^{2}+m^2-A[\chi_{r}-\bar
q_{r}]\label{mr},
\end{equation}
where $\chi(\omega_r)$ and $q^{\alpha\beta}(\omega_r) $ are,
respectively, the Fourier components of
 $\chi(\tau-\tau')$ and
$q^{\alpha\beta}(\tau-\tau')$. We use for these, the simplified
notation $\chi_r\equiv \chi(\omega_{r})$ and $
q^{\alpha\beta}_r\equiv q^{\alpha\beta}(\omega_{r})$.

From (\ref{hamiltonianoefetivo1}), we can show that
\begin{equation}
Q_{i}^{\alpha\beta}(\tau,\tau') = \langle \hat S_i^\alpha(\tau) \hat
S_i^\beta(\tau') \rangle
\end{equation}
and therefore, according to the previous decomposition of $Q$ into
$\chi$'s and $q$'s, we can identify $\chi_0 $ as the static magnetic
susceptibility, whereas the integrated susceptibility is given by
\begin{equation}
\chi_\mathrm{I}= \sum_r \chi_{r} .
\end{equation}
The EA order parameter, used to detect the SG phase \cite{ea,by},
accordingly, is given by
\begin{equation}
 q_{\mathrm{EA}} = T \bar q_0
\end{equation}

where $\bar q_0 $ is defined in (\ref{qbar}).

\subsection{The Stationary-Phase Equations}

By taking the variations of $\bar f$ with respect to the variables
$\sigma^\alpha, m^2, \chi_r, q^{\alpha\beta}_r$, respectively, we
obtain the stationary-phase equations (SPE), which are listed below.


%
\begin{equation}
\frac{1}{n}\left[\left[m^2- A\chi_0\right]\delta^{\alpha\beta}- A q^{\alpha\beta}_0\right]\sigma^\beta=0
\label{I}
\end{equation}
\begin{equation}
\sigma^{2}={\rho}_{s}-\frac{T}{2\pi}\sum_{\omega_{r}}\ln\left(1+\frac{\Lambda^{2}}{M_{r}}\right)+
2A\sum_{\omega_{r}}\bar q_{r}G_{r}
\label{II}
\end{equation}
\begin{equation}
3DT \chi(-\omega_{r})=\frac{TA}{4\pi}\ln\left(1+\frac{\Lambda^{2}}{M_{r}}\right)+ A^2\bar q_{r}G_{r}
- A\sigma^{2}\delta_{r0}
\label{III}
\end{equation}
\begin{equation}
3DT q^{\alpha\beta}(-\omega_r) = A^2 \bar q(\omega_{r})G_r+ A \frac{\sigma_{\alpha}\sigma_{\beta}}{2n}\delta_{r0}
\label{IV}
\end{equation}
where
\begin{equation}
G_r=\frac{T}{4\pi}\left[\frac{1}{M_r}-\frac{1}{\Lambda^2+M_r}\right ]\label{gr}
\end{equation}
 and $\Lambda =1/a$ is the high-momentum cutoff.

Observe that in the absence of disorder ($D,A\rightarrow 0$) (\ref{III}) and (\ref{IV}) disappear and (\ref{I}), (\ref{II})
reduce to the well-known corresponding equations for the continuum limit of the pure Heisenberg model \cite{sach,chn,em}. We
henceforth will only consider the case $D\neq 0$.

Equation (\ref{IV}) tells us that in our approximation all the
$q^{\alpha\beta}$'s are equal, whenever the $\sigma^{\alpha}$'s
vanish.

\section{The Phase Diagram}

\setcounter{equation}{0}

\subsection{Paramagnetic and Spin-Glass Phases}

\subsubsection{Preliminaries}

We start by searching for PM and SG phases. In both of them we have $\sigma =0$.
Considering this fact and summing (\ref{IV}) in $\alpha, \beta$, yields
\begin{equation}
 \bar q(-\omega_r) \Gamma= \bar q(\omega_{r})G_r,\label{IVa}
\end{equation}
where $\Gamma=\frac{T\gamma}{4\pi\Lambda^2}$ and
\begin{equation}
\gamma=\frac{3\pi\rho_s^2\Lambda^2}{D}= 3\pi \left(\frac{\bar J}{\Delta J}  \right )^2.\label{gama}
\end{equation}

Inserting (\ref{IVa}) in (\ref{III}), we get
\begin{equation}
m^2 + \omega_r^2 = \frac{\Lambda^2}{  e^{6\pi\rho_s (\chi_{-r}-\bar q_{-r})}-1 } + A (\chi_r-\bar q_r),
\label{m2}
\end{equation}
or equivalently
\begin{equation}
M_r = \frac{\Lambda^2}{  e^{6\pi\rho_s (\chi_{-r}-\bar q_{-r})}-1 }
\label{M}
\end{equation}

\begin{figure}[ht]
\centerline {
\includegraphics
[clip,width=0.5\textwidth ,angle=0 ] {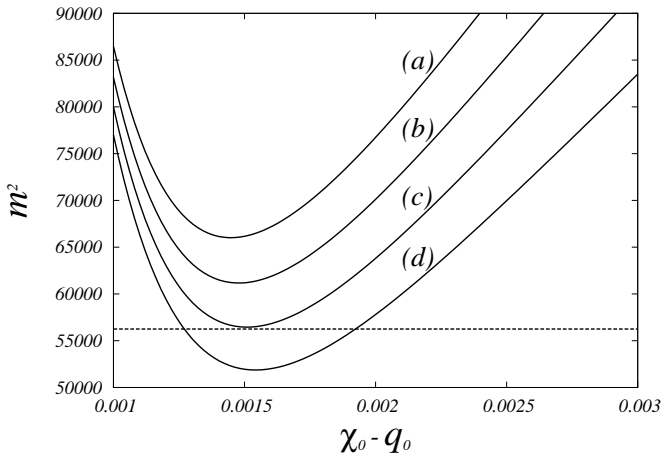} }
\caption{Function appearing in the rhs of (\ref{m2}) ($r=0$) for
different values of the Gaussian width $\Delta J$ (a: 220K, b: 210K,
c: 200K, d: 190K). The horizontal dashed line represents $m^2$. A PM
phase will only occur when this line intercepts the function ($m^2 >
m^2_0$). The physical solution corresponds to the left branch. $m$
is in $K$ and $\chi_0$ and $q_0$ are in $K^{-1}$ }\label{massad}
\end{figure}

We can use (\ref{m2}) in order to determine $\chi_r-\bar q_r$. Let
us start with $r =0$. In this case, both $\chi_0$ and $\bar q_0$ are
real and equation (\ref{m2}) is depicted in Fig. \ref{massad}.

The function on the r.h.s. of (\ref{m2}) has a minimum at
\begin{equation}
\chi_0 = \bar\chi_{\mathrm {cr}} = \frac{1}{6\pi \rho_s}\ln\left [1+\gamma +\frac{\gamma}{2}\left [\left ( 1 + \frac{4}{\gamma}\right )^{1/2}-1 \right ]\right ]\label{chicr},
\end{equation}
at which the function has the value
$$
m_0^2=\frac{\Lambda^2}{\gamma}\left[\frac{1}{1+\frac{1}{2}\left[\left ( 1 + \frac{4}{\gamma}\right )^{1/2}-1 \right]}\  +\right.
$$
\begin{equation}
\left.
+\ln\left [1+\gamma +\frac{\gamma}{2}\left [\left ( 1 + \frac{4}{\gamma}\right )^{1/2}-1 \right ]\right ]\right ].
\end{equation}
It follows that for $r = 0$
(\ref{m2}) will only have solutions for $m^2 > m_0^2$.
In this case, however, these solutions of (\ref{m2})
are clearly not compatible with the existence of nontrivial solutions ($\bar q_0 \neq 0$) of (\ref{IVa}).

We can see this as follows. A solution $\bar q_0 \neq 0$ of (\ref{IVa})
would imply $G_0=\Gamma$.  In the range of values of $m^2$ for which
(\ref{m2}) has solutions ($m^2  > m_0^2$), however these will be such that $G_0<\Gamma$. This is so because $G_0$ is a monotonically decreasing function of $M_0$ and according to (\ref{M}), $M_0$ is a monotonically increasing function of $m^2$, such that precisely $G_0(M_0(m_0^2))= \Gamma$, as
can be immediately inferred from (\ref{gr}) and from
\begin{equation}
M_0(m_0^2)=\frac{\Lambda^2}{2}\left[\left ( 1 + \frac{4}{\gamma}\right )^{1/2}-1 \right].
\label{Mzero}
\end{equation}
It immediately follows that, for $m^2  > m_0^2$, we will have
$G_0(M_0(m^2))< \Gamma$.
 Hence, the only possible solution of (\ref{IVa}), for $m^2  > m_0^2$,
is $\bar q_0 = 0$.
For $m^2 < m_0^2 $, however, (\ref{m2}) no longer provides a solution for $\chi_0$, hence we may now have $\bar q_0 \neq 0$. The static susceptibility
$\chi_0$ is now determined by (\ref{IVa}), namely, $G_0=\Gamma$.

Now consider the case $r \neq 0 $. We show in Appendix A that we
always have $q_r = 0$ ( for $r \neq 0$), because otherwise
(\ref{m2}) and (\ref{IVa}) again become incompatible. We will always
have, therefore, the $\chi_{r\neq 0}$ determined by (\ref{m2}).

Since $q_{\mathrm{EA}} = T \bar q_0$  it follows that $\bar q_0$ is also a SG order parameter and
we conclude that the former phase ($m^2 > m_0^2$) is a paramagnetic phase $(\sigma = 0, q_{\mathrm{EA}} = 0)$, whereas the latter ($m^2 < m_0^2$) is a SG phase $(\sigma = 0, q_{\mathrm{EA}} \neq 0)$.
The phase transition occurs at $m^2 = m_0^2$.

The ratio $\frac{1}{\gamma}$, which appears in the expression of the critical mass $m_0$, is a measure of the amount of frustration in the system, as we can infer from (\ref{gama}) and the actual perturbation parameter.
Since we are working in the regime of weak disorder, we take $\gamma \gg 1$. In the unperturbed limit where the disorder is removed $(\gamma \rightarrow\infty)$, we would have $m_0^2=0$ and the SG phase would no longer exist.
 We see again that a disorder perturbation would be impossible in the original EA model, where $\gamma=0$.

\subsubsection{The Paramagnetic Phase}

Let us now use the SPE in order to derive, expressions for the
susceptibilities $\chi_\mathrm{I}$ and $\chi_0$ in the PM phase,
where $m^2 > m_0^2$. Inserting (\ref{III}) in (\ref{II}), for
$\sigma =0$ and $\bar q_0 =0$ (and also $\bar q_{r\neq 0}=0$), we
readily obtain, for the integrated susceptibility
\begin{equation}
\chi_\mathrm{I}^{\mathrm{PM}}= \frac{1}{3T}\label{chiipm} .
\end{equation}

It follows from (\ref{III}) that
\begin{equation}
\chi_0^{\mathrm{PM}}= \frac{1}{3T}-\frac{1}{6\pi \rho_s} \sum_{\omega_{r}\neq 0}[\ln(\Lambda^2+M_r)-\ln M_r] .
\end{equation}

The previous sums are dominated by large values of $\omega_r$. In
this case, we show in Appendix A that (\ref{m2}) or equivalently
(\ref{mf1}) and (\ref{mf2}) yield the solution
\begin{equation}
\chi(\omega_{r})\simeq \frac{\Lambda^2}{6\pi \rho_s\left(m^2+\omega_{r}^2\right)}.
\label{xir}
\end{equation}
We may then evaluate the two sums above, obtaining for them an explicit expression
\begin{equation}
\sum_{\omega_{r}\neq 0}[\ln(\Lambda^2+M_r)-\ln M_r] = \ln
\Upsilon(m^2,T,\gamma)\label{somaY},
\end{equation}
where
$$
 \Upsilon=\frac{\sinh^{2}\left(\frac{\sqrt{X_{+}}}{2T}\right)
\sinh^{2}\left(\frac{\sqrt{X_{-}}}{2T}\right)}
{\sinh^{2}\left(\frac{1}{2T}\sqrt{m^{2}+\frac{\Lambda^{2}}{\sqrt{\gamma}}}\right)
\sinh^{2}\left(\frac{1}{2T}\sqrt{m^{2}-\frac{\Lambda^{2}}{\sqrt{\gamma}}}\right)}
$$
\begin{equation}
\times \frac{\left(m^{2}+\frac{\Lambda^{2}}{\sqrt{\gamma}}\right)\left(m^{2}-\frac{\Lambda^{2}}
{\sqrt{\gamma}}\right)}{X_{+}X_{-}}\label{Y},
\end{equation}
with
\begin{equation}
X_{\pm}=(\Lambda^{2}+2m^{2})\left[\frac{1}{2}\mp\frac{1}{2}\sqrt{\frac{\Lambda^{4}}
{(\Lambda^{2}+2m^2)^2}\left[1+\frac{4}{\gamma}\right]
}\right].
\end{equation}

The static susceptibility in the PM phase, therefore, is given by
\begin{equation}
\chi_0^{\mathrm{PM}}= \frac{1}{3T}- \frac{1}{6\pi \rho_s} \ln
\Upsilon(m^2,T,\gamma) \label{chi0pm}.
\end{equation}

The function $\Upsilon(m^2,T,\gamma)$ has the properties
\begin{equation}
\Upsilon(m^2,T,\gamma)\stackrel{T>>\Lambda}\longrightarrow 1
\end{equation}
and
\begin{equation}
\ln \Upsilon(m^2,T,\gamma)\stackrel{T\rightarrow 0}\longrightarrow
\frac{2\pi}{T}\rho_0,
\end{equation}
where (for $\gamma >> 1$)
\begin{equation}
\rho_0=\frac{\Lambda}{2\pi}\left[1+\frac{1}{\gamma}\left[1+\frac{1}{2}\ln(1+\gamma)\right]\right].
\label{r0}
\end{equation}

We also have
\begin{equation}
\frac{1}{3T}- \frac{1}{6\pi \rho_s} \ln \Upsilon(m^2,T,\gamma)
\stackrel{m^2 \rightarrow m_0^2}\longrightarrow
\bar\chi_{\mathrm{cr}},
\end{equation}
where $\bar\chi_{\mathrm{cr}}$ is given by (\ref{chicr}), implying that
 the critical value of  $\chi_0^{\mathrm{PM}}$ is
  $\bar\chi_{\mathrm{cr}} $.

We see that $\chi_0^{\mathrm{PM}}$ satisfies the Curie law at high-temperatures
\begin{equation}
\chi_0^{\mathrm{PM}} \stackrel{T>>\Lambda}\longrightarrow \frac{1}{3T}
\end{equation}
and diverges as
\begin{equation}
\chi_0^{\mathrm{PM}} \stackrel{T\rightarrow 0}\longrightarrow \frac{1}{3T}\left[1-\frac{\rho_0}{\rho_s}\right],
\end{equation}
for $T\rightarrow 0$. As we will see this is the expected behavior
for $(\rho_s >\rho_0)$, where an AF phase appears at $T=0$. For $(\rho_s <\rho_0)$, conversely, we will
 see that the PM-SG phase transition occurs at a finite $T_c$ (Fig. \ref{FigRhoxTc} and Fig. \ref{rigidez})
and the previous expression is no longer valid.

Notice that the following general relation involving the integrated susceptibility \cite{by}
is automatically satisfied by $\chi_\mathrm{I}^{\mathrm{PM}}$ and $\bar q^{\mathrm{PM}}(\omega_{r}) = 0$:
\begin{equation}
\chi_\mathrm{I}=\frac{1}{3T}-\frac{1}{3}\sum_{\omega_{r}}\bar
q(\omega_{r}) \label{ggr}
\end{equation}

\subsubsection{The Spin-Glass Phase}

We now turn to the the SG phase ($m^2 < m_0^2$).
Now (\ref{IVa}), or equivalently  $G_0=\Gamma$ implies
\begin{equation}
M_0 = \frac{\Lambda^2}{2}\left[ \left (1+ \frac{4}{\gamma}\right )^{1/2}-1 \right ] = \frac{\Lambda^2}{\gamma} \left[ 1+ \mathrm{O}\left( \frac{1}{\gamma}\right ) \right ]\label{M0},
\end{equation}
which coincides with (\ref{Mzero}).

Then (\ref{II}) yields (for $\gamma >> 1$)
\begin{equation}
\bar q_0^{\mathrm{SG}}= \frac{1}{3T}-\frac{1}{6\pi \rho_s}\ln
\Upsilon(m^2,T,\gamma)-\frac{1}{6\pi \rho_s}\ln (1+\gamma).
\label{q0sg}
\end{equation}
From this and (\ref{M0}) we obtain
\begin{equation}
\chi_0^{\mathrm{SG}}= \frac{1}{3T}-\frac{1}{6\pi \rho_s}\ln
\Upsilon(m^2,T,\gamma)-\frac{\rho_s}{2D}[m_0^2-m^2]. \label{x0sg}
 \end{equation}

Since (\ref{III}) still holds for $r\neq 0$, we still have
\begin{equation}
 \sum_{\omega_r\neq 0}\chi(\omega_{r}) = \frac{1}{6\pi \rho_s}\ln  \Upsilon(m^2,T,\gamma)
 \end{equation}
 and therefore we get
\begin{equation}
\chi_\mathrm{I}^{\mathrm{SG}}=\frac{1}{3T}-\frac{\rho_s}{2D}[m_0^2-m^2],
\label{xisg}
 \end{equation}

Comparing (\ref{x0sg}) and (\ref{chi0pm}) and also (\ref{xisg}) and (\ref{chiipm}) we can identify a clear cusp at the transition
appearing in both susceptibilities.
This is an important result, since
the presence of
these cusps is a benchmark of the SG transition and
has been experimentally observed in many materials presenting a SG phase \cite{by}.

\subsection{The N\'eel Phase}

Let us now search for an ordered N\'eel phase, for which $\sigma \neq 0$. We see that in this case the quantity
between brackets in (\ref{I}) must vanish. By summing it in $\alpha, \beta$, we conclude that $M_0 =0$ in this phase. According to
(\ref{II}), however, this can only happen at $T=0$, otherwise we would have an unphysical infinite imaginary
 staggered magnetization $\sigma$.
This is in agreement with the Mermin-Wagner theorem \cite{mw} and is
a clear evidence that our approach goes beyond mean-field.

Now,  $M_0 =0$ implies
\begin{equation}
\chi_0^{\mathrm{AF}} - \bar q_0^{\mathrm{AF}} = \frac{m^2}{A}.
\label{chiq}
\end{equation}

On the other hand, for  $\sigma \neq 0$, (\ref{II}) and (\ref{III}) imply, instead of (\ref{chiipm})
\begin{equation}
\chi_\mathrm{I}^{\mathrm{AF}}  = \frac{1}{3T}\left(1 - \frac{2\sigma^2}{\rho_s}  \right ).
\label{chiiaf}
\end{equation}
From this, it follows that
\begin{equation}
\chi_0^{\mathrm{AF}}= \frac{1}{3T}- \frac{1}{6\pi \rho_s} \ln
\Upsilon(m^2,T,\gamma) - \frac{2\sigma^2}{3T\rho_s} \label{chi0af},
\end{equation}
which in the limit $T\rightarrow 0$ reduces to
\begin{equation}
\chi_0^{\mathrm{AF}} \stackrel{T\rightarrow 0}\sim \frac{1}{3T} \left[1-\frac{\rho_0}{\rho_s}\right]
- \frac{2\sigma^2}{3T\rho_s} \label{chi0af1},
\end{equation}
for $\rho_s > \rho_0$.

On the other hand, (\ref{ggr}) leads to
\begin{equation}
\bar q_0^{\mathrm{AF}}  = \frac{1}{T}\left( \frac{2\sigma^2}{\rho_s}  \right ).
\label{q0af}
\end{equation}

Equations (\ref{chiq})(\ref{chi0af1}) and (\ref{q0af}), allow us to solve for $\chi_0^{\mathrm{AF}}$, $\bar q_0^{\mathrm{AF}}$
and $\sigma$. We get, for $T\rightarrow 0$,
\begin{equation}
\sigma^2 = \frac{1}{8} \left[\rho_s-\rho_0\right]
 \label{saf}
\end{equation}
and
\begin{equation}
\chi_0^{\mathrm{AF}}= \bar q_0^{\mathrm{AF}} =\frac{1}{4T\rho_s} \left[\rho_s-\rho_0\right] \label{chiqaf}.
\end{equation}

This implies, according to (\ref{chiq}) a zero spin gap: $m^2=0$.

From (\ref{chiiaf}) and (\ref{saf}) we obtain
\begin{equation}
\chi_\mathrm{I}^{\mathrm{AF}}  = \frac{1}{12T}\left[3 + \frac{\rho_0}{\rho_s}  \right ].
\label{chiiaf1}
\end{equation}

Notice that the susceptibilities diverge for $T\rightarrow 0$ as they should. The EA parameter, however, remains finite:
\begin{equation}
\bar q_{\mathrm{EA}}^{\mathrm{AF}}  = \frac{1}{4\rho_s} \left[\rho_s-\rho_0\right].
\label{qEAaf}
\end{equation}

We see that, indeed, there is an AF phase characterized by $(\sigma \neq 0, q_{\mathrm{EA}} \neq 0)$ on the line $( T=0, \rho_s>\rho_0)$,
 with
$\rho_0$ given by (\ref{r0}).
In the absence of disorder $(\Delta J=0, \gamma \rightarrow\infty)$,
$\rho_0\rightarrow \rho_0(0)=\frac{\Lambda}{2\pi}$, which is the well-known quantum critical coupling determining the boundary of the
AF phase in the pure 2D AF Heisenberg model at $T=0$ \cite{chn,sach,em}. The effect of disorder on the AF phase is to displace the
 quantum critical point (QCP) to the right.
This result should be expected on physical grounds: in the presence of disorder a larger coupling is required, to stabilize an
ordered AF phase.

\subsection{Critical Curve}

We have seen that the parameter $m^2$ determines the transition
between the PM and SG phases. It is important, consequently, to see
how it depends on the control parameters of our system, namely, $T$,
$\bar J$ and $\Delta J$, or, equivalently, $T$, $\rho_s$ and
$\gamma$. We may obtain an equation for $m^2$ by using (\ref{ggr}),
(\ref{q0sg}), (\ref{xisg}) and the fact that $\bar q_{r\neq 0}=0$.
These yield
\begin{equation}
m^2-m_0^2=\frac{2D}{3\rho_s}\left[\bar\chi_{\mathrm {cr}}
 - \frac{1}{3T}+\frac{1}{6\pi \rho_s}\ln \Upsilon(m^2,T,\gamma)\right]\label{deltaeme}.
\end{equation}

From this we get
\begin{equation}
\bar q_0^{\mathrm{SG}}=\frac{3\rho_s}{2D}\left[m_0^2-m^2\right]
\end{equation}
 for $m^2 < m_0^2$.
We see that $\bar q_0^{\mathrm{SG}} \rightarrow 0$ at the transition as it should.
For $m^2 > m_0^2$ we have $\bar q_0^{\mathrm{PM}}=0$ as seen above.

We may determine the critical curve by observing that the critical
condition $m^2=m_0^2$ implies
\begin{equation}
 \frac{1}{3T_c}-\frac{1}{6\pi \rho_s}\ln \Upsilon(m_0^2,T_c,\gamma)= \bar\chi_{\mathrm {cr}}
\end{equation}

\begin{figure}[ht]
\centerline {
\includegraphics
[clip,width=0.5\textwidth ,angle=0 ] {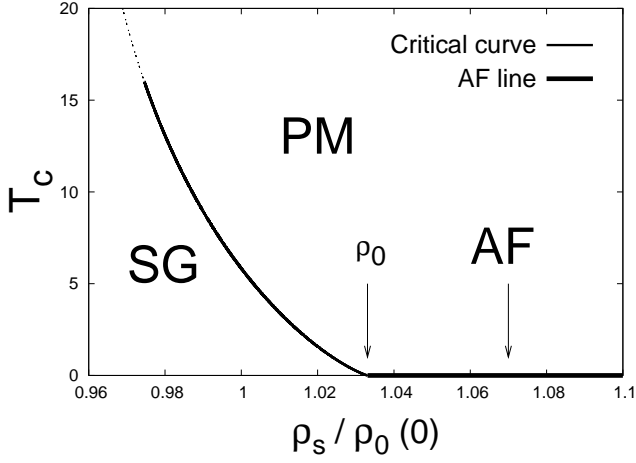} } \caption{Phase
diagram for a fixed value of $\gamma$. ($\gamma=10^2,\Lambda=10^3$).
The critical curve corresponds to (\ref{ccritica}) and is valid near
the QCP $\rho_0$ (solid curve). $\rho_0(0)=\Lambda/2\pi$ is the QCP
of the pure AF system. Notice that disorder besides creating the SG
phase, displaces the QCP to the right. The value ascribed to
$\Lambda$ is a realistic one in $K$ ($\Lambda \rightarrow
\frac{\hbar v_s}{k_{B}}\Lambda$; $v_s$: spin-wave velocity). The
resulting temperatures naturally appear with the correct order of
magnitude, in $K$, found in real SG systems \cite{by}.}
\label{FigRhoxTc}
\end{figure}

For $T_c\ll\Lambda$, which corresponds the situation found in realistic systems, this becomes, near the
quantum critical point ($\rho_s \lesssim \rho_0$),
\begin{equation}
\frac{T_c}{2\pi}\left[ \ln\left( \frac{\Lambda}{T_c}\right)^2-\ln(1+\gamma)\right]=\rho_0-\rho_s\label{ccritica},
\end{equation}
which is the equation for the critical curve separating the PM and
SG phases. Notice that it meets the $T=0$ axis, precisely at the
quantum critical point $\rho_0$, separating the SG from the AF
phase. We plot the $T_c \times \rho_s$ phase diagram corresponding
to (\ref{ccritica}), for a fixed value of $\gamma$ in
Fig.\ref{FigRhoxTc}.

\begin{figure}[ht]
\centerline {
\includegraphics
[clip,width=0.5\textwidth ,angle=0 ] {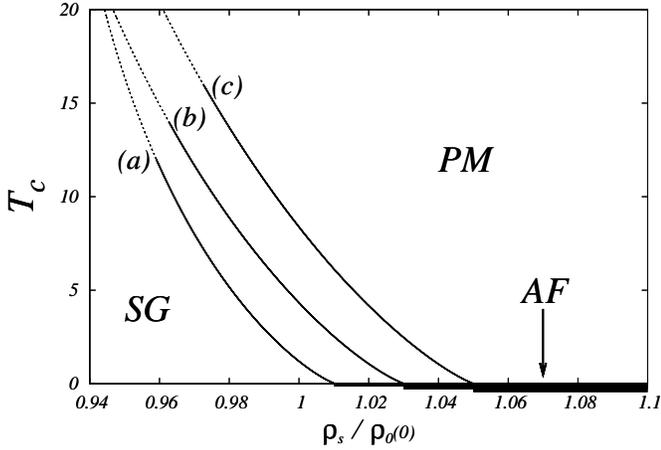} } \caption{Phase
diagram for $\Lambda=10^3$ and $\Delta J$ (a: 200K, b: 350K, c:
460K). Notice that the AF phase (thick line) is displaced to the
right as we increase the disorder.}\label{rigidez}
\end{figure}

 In Fig. \ref{rigidez}, we plot again  $T_c
\times \rho_s$, for a fixed value of the amount of disorder, namely,
the Gaussian width $\Delta J$.

\begin{figure}[ht]
\centerline {
\includegraphics
[clip,width=0.5\textwidth ,angle=0 ] {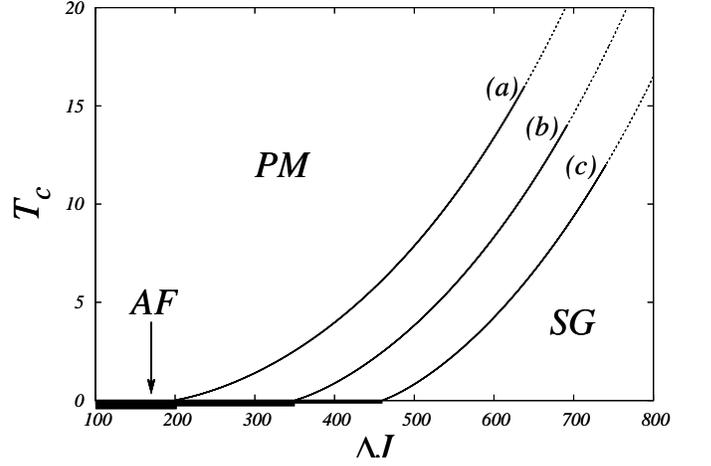} } \caption{Phase
diagram $T_c \times \Delta J$, for $\Lambda=10^3$ and
$\rho_s/\rho_0(0)$ (a: 1.01, b: 1.03, c: 1.05). Notice that as we
increase the value of the center of the Gaussian ($\rho_s$), a
larger amount of disorder will be required for the SG
phase.}\label{desordem}
\end{figure}

Now we plot in Fig. \ref{desordem} the associated $T_c \times \Delta
J$ phase diagram, for different values of the spin stiffness. The
latter shows the AF-SG transition as a function of increasing
disorder. This is the type of transition which is observed in the
high-Tc cuprates. In order to describe it, we must relate the doping
parameter of these materials to our disorder parameter, $\Delta J$.
We are presently investigating this point.

Below, we plot the quantum critical point $\rho_0$ as a function of
the amount of disorder, for different values of $\rho_s$.
\begin{figure}[ht]
\centerline {
\includegraphics
[clip,width=0.5\textwidth ,angle=0 ] {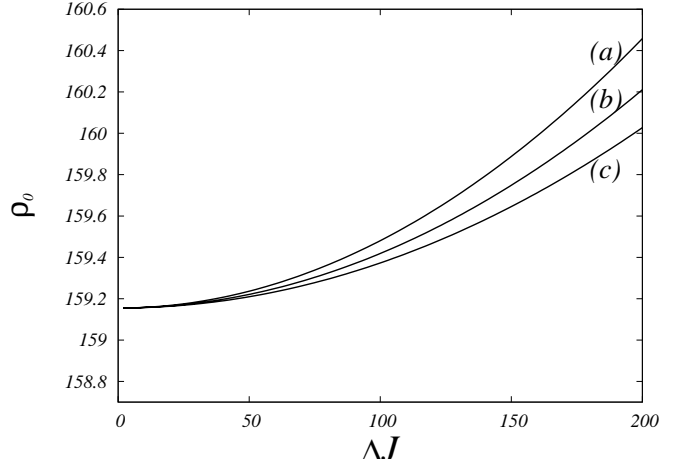} }
\caption{The quantum critical point as a function of the amount of
disorder for different values of the average coupling, $\rho_s$( a:
180K , b: 200K, c: 220K ). Notice that in the limit of zero disorder
$\rho_0 \rightarrow \rho_0(0)= \Lambda/2\pi$.} \label{rho0}
\end{figure}

\subsection{Critical Behavior}

The critical behavior of relevant quantities may be determined by
analyzing the function $\Upsilon(m^2,T,\gamma)$ for $T\sim T_c$ and
$m^2 \sim m_0^2$. This yields, near the transition, for $\rho_s
\lesssim \rho_0$,
\begin{equation}
\frac{1}{3T}-\frac{1}{6\pi \rho_s}\ln \Upsilon(m^2,T,\gamma) \sim
\left(\frac{T_c}{T}\right) \bar\chi_{\mathrm{cr}}
\end{equation}
and
\begin{equation}
 m^2-m_0^2 \sim 4\pi\Lambda \left[\frac{T-T_c}{T_c}\right][\rho_0-\rho_s].
\label{mm0}
\end{equation}
From these expressions and (\ref{chi0pm}), (\ref{chiipm}), (\ref{x0sg}), (\ref{xisg}) and (\ref{q0sg}) we can fully determine the critical behavior
of the SG order parameter and susceptibilities  for $T\gtrsim T_c$ and $\rho_s < \rho_0$ :

\begin{equation}
\chi_0^{\mathrm{PM}}\sim \left(\frac{T_c}{T}\right) \bar\chi_{\mathrm{cr}}\ \ ;\ \chi_\mathrm{I}^{\mathrm{PM}}=\frac{1}{3T}\ \ ;\ \bar q_0^{\mathrm{PM}}=0
\label{chipm},
\end{equation}
and for $T\lesssim T_c$ and $\rho_s < \rho_0$:
\begin{equation}
\chi_0^{\mathrm{SG}}\sim \left(\frac{T_c}{T}\right) \bar\chi_{\mathrm{cr}} -\frac{2\pi\Lambda\rho_s}{D}\left [\frac{T_c-T}{T_c}\right][\rho_0-\rho_s]
\label{chisg}
\end{equation}
\begin{equation}
\chi_\mathrm{I}^{\mathrm{SG}}\sim \frac{1}{3T} -\frac{2\pi\Lambda\rho_s}{D}\left [\frac{T_c-T}{T_c}\right][\rho_0-\rho_s]
\label{chisg},
\end{equation}
\begin{equation}
\bar q_0^{\mathrm{SG}}\sim \frac{6\pi\Lambda\rho_s}{D}\left [\frac{T_c-T}{T_c}\right][\rho_0-\rho_s]
\label{qsg},
\end{equation}
follows.

\begin{figure}[ht]
\centerline {
\includegraphics
[clip,width=0.5\textwidth
 ,angle=0
] {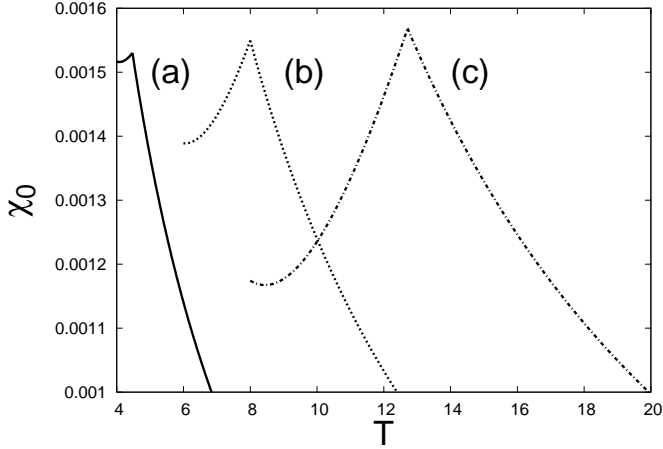} }
\caption{The static susceptibility for different values of $\rho_s$:
$\rho_s/\rho_0(0)= $(a)1.005 ,(b)0.993, (c)0.981.
 The cusps, characteristic of the SG transition, occur at the corresponding critical temperatures
(in $K$). $\chi_0$ is in $K^{-1}$.}
\label{Chi0-1}
\end{figure}

We plot $\chi_0$ in Fig. \ref{Chi0-1}. We can see the characteristic
cusps of the SG transition occurring in these magnetic
susceptibilities.

For $\rho_s>\rho_0$ and $T>0$, we always have $m^2>m_0^2$ and $\sigma=\bar q_0=0$ i.e. the system is in the PM phase for any
finite temperature.

\subsection{Dependence on Disorder}

Let us now examine the $\Delta J$ dependence of the phase diagram. For this we make $\Delta J\rightarrow \Delta J(1+ \epsilon)$, with $|\epsilon| \ll 1$, for fixed $\rho_s$, and study how the
relevant quantities change. We find,
\begin{equation}
\rho_0^\epsilon-\rho_0 =  \epsilon\frac{\Lambda}{2\pi\gamma}\ln\gamma.
\label{depdes}
\end{equation}
For $\rho_s<\rho_0$ and a fixed $\rho_s$, we also obtain
\begin{equation}
 \frac{T_c^\epsilon-T_c}{T_c}=
\frac{[m_0^2]^\epsilon-m_0^2}{4\pi\Lambda\rho_0} = \frac{\rho_0^\epsilon-\rho_0}{\Delta\rho}  =
\epsilon\frac{\Lambda}{2\pi\gamma\Delta\rho}\ln\gamma.
\label{depdes},
\end{equation}
where $\Delta\rho=\rho_0-\rho_s$ .

The SG order parameter changes as
\begin{equation}
[\bar q_0^{\mathrm{SG}}]^\epsilon-\bar q_0^{\mathrm{SG}}= \frac{\epsilon}{\pi\rho_s}\ln\gamma .
\end{equation}
We see that increasing the amount of disorder, through an increment of the Gaussian width,
will increase $\rho_0$ and, for a fixed $\rho_s$, also $T_c$, $m_0^2$ and $\bar q_0^{\mathrm{SG}}$. Conversely, decreasing the amount of disorder by narrowing the Gaussian width will produce the opposite effects.


\subsection{ Effect of Quantum Fluctuations}

As mentioned before, the effect of quantum fluctuations, introduced
by integration over the quadratic fluctuations of the CP$^1$ fields
and embodied in the last term of (\ref{f}) is essential for the
obtainment of a sensible solution for this system. Should we do a
pure mean-field approach, by disregarding these fluctuations, we
would obtain a N\'eel state for any temperature, in obvious
disagreement with the Mermin-Wagner theorem. Conversely, taking
these fluctuations into account washes out the N\'eel phase to
$T=0$, but leaving a spin glass state below the critical curve. The
fact that the AF phase is removed to T=0 shows that our approach
transcends the mean-field approximation. The SG phase is robust to
such fluctuations. In the spirit of the loop expansion that is being
done, we assume that higher quantum fluctuations will not change
this picture qualitatively. Anyway, considering only the quadratic
fluctuations is always a valid first approach to a difficult
problem.

\section{The Thermodynamic Stability}

\setcounter{equation}{0}

We finally consider the important question of the thermodynamic
stability of the phases. As is well known the replica symmetric
mean-field solution of the SK model turned out to be unstable. It
is, therefore, absolutely necessary to analyze the stability of any
solution to a SG. We will focus on the SG and PM phases. The
stability of the AF phase should not be a problem.

For studying the stability of the solution, we must consider the Hessian matrix of the free-energy
density
\begin{equation}
 \bar f = \bar f[\sigma^\alpha; q^{\alpha\beta}(\omega_0),...q^{\alpha\beta}(\omega_r)...;
 \lambda;\chi(\omega_0),..., \chi(\omega_r),...] ,
\end{equation}
where the variables $q^{\alpha\beta}(\omega_r)$ and $\chi(\omega_r)$
are complex, such that $\chi_{-r} = \chi^*_r$ and
$q^{\alpha\beta}_{-r}= q^{*\ \alpha\beta}_r$.

The Hessian is given by
\begin{equation}
\mathbb{H}_{ij} = \frac{\partial^2 \bar{f}}{\partial \phi_i \partial
\phi_j}. \label{hessian}
\end{equation}
where $\phi_i = \sigma^\alpha, q^{\alpha\beta}(\omega_0),
\mathrm{Re}\ q^{\alpha\beta}(\omega_r), \mathrm{Im}\
q^{\alpha\beta}(\omega_r), \lambda,\chi(\omega_0),$ $ \mathrm{Re}\
\chi(\omega_r), \mathrm{Im}\ \chi(\omega_r)$, where the index $r$
runs from $1$ to $\infty$.
 This is a matrix with entries of dimensions
$\lim_{n\rightarrow
0}[n;n(n-1)_0,...,2n(n-1)_r,...;1;1_0,...,2_r,...]$ corresponding,
respectively, to derivatives with respect to each of the above
variables. Since we are interested here in the SG and PM phases, we
will take (\ref{hessian}) at $\sigma =0$.

The elements of the Hessian matrix are as follows: there are four
overall diagonal and six crossed elements, namely $\sigma\sigma$,
$qq$, $\lambda\lambda$, $\chi\chi$, $\sigma q$, $\sigma\lambda$,
$\sigma\chi$, $q\lambda$, $q\chi$ and $\lambda\chi$.

The $\sigma\sigma$ term is,
\begin{equation}
\frac{\partial^{2}\bar f}{\partial\sigma^{\alpha}\partial\sigma^{\beta}}=\frac{1}{n}\left [ M_{0}\delta^{\alpha\beta}
-A \bar{q}_{0} \mathbb{C}^{\alpha\beta} \right ]
\label{ss}
\end{equation}

The $q_r q_s$ terms are proportional to $\delta_{rs}$. For $r=0$
\begin{equation}
\frac{\partial^{2}\bar f}{\partial q^{\alpha\beta}_{0}\partial
q^{\gamma\eta}_{0}}=\left[a_{0}\delta^{\alpha\gamma}
\delta^{\beta\eta}+b_{0}
\mathbb{C}^{\alpha\gamma}\mathbb{C}^{\beta\eta}\right]  \equiv K_0 ,
\end{equation}

where
\begin{equation}
 a_{r}=A^{2}\frac{\Gamma}{n}\ \ \ ;\ \ \  b_{r}=A^{2}\frac{G_{r}-H_{r}}{[n(n-1)]^{2}}
 \label{ab}
\end{equation}
and

\begin{equation}
H_{r}=A\frac{T}{4\pi}\bar{q}_{r}\left[\frac{1}{M_{r}^2}-\frac{1}{(\Lambda^{2}+M_{r})^{2}}\right].
\end{equation}

For the $r \neq 0$ blocks, we have four terms, namely

$$
\frac{\partial^{2}\bar f}{\partial \mathrm{Re}\
q^{\alpha\beta}(\omega_r)\partial \mathrm{Re}\
q^{\alpha\beta}(\omega_s)}=
$$
\begin{equation}
\left[a_{r}\delta^{\alpha\gamma} \delta^{\beta\eta}+ \mathrm{Re}\
b_{r} \mathbb{C}^{\alpha\gamma}\mathbb{C}^{\beta\eta}\right]  \equiv
K_{r}(11) \delta_{rs} \label{k1},
\end{equation}

$$
\frac{\partial^{2}\bar f}{\partial \mathrm{Im}\
q^{\alpha\beta}(\omega_r)\partial \mathrm{Im}\
q^{\alpha\beta}(\omega_s)}=
$$
\begin{equation}
\left[a_{r}\delta^{\alpha\gamma} \delta^{\beta\eta}- \mathrm{Re}\
b_{r} \mathbb{C}^{\alpha\gamma}\mathbb{C}^{\beta\eta}\right]  \equiv
K_{r}(22) \delta_{rs} \label{k2},
\end{equation}

$$
\frac{\partial^{2}\bar f}{\partial \mathrm{Re}\
q^{\alpha\beta}(\omega_r)\partial \mathrm{Im}\
q^{\alpha\beta}(\omega_s)}=
$$
\begin{equation}
\left[- \mathrm{Im}\ b_{r}
\mathbb{C}^{\alpha\gamma}\mathbb{C}^{\beta\eta}\right]  \equiv
K_{r}(12) \delta_{rs} = K_{r}(21) \delta_{rs}\label{k3} .
\end{equation}

The $\lambda\lambda$ term is,

\begin{equation}\frac{\partial^{2}\bar f}{\partial \lambda\partial
\lambda} =
- \frac{\partial^{2}\bar f}{\partial m^{2}\partial
m^{2}}=\sum_{r}[G_{r}+H_{r}]\equiv \varphi
\label{fi}
\end{equation}

The $\chi_r \chi_s$ terms are also proportional to $\delta_{rs}$.
For $r=0$, we have
\begin{equation}
\frac{\partial^{2}\bar f}{\partial\chi_{0}\partial
\chi_{0}}=A^{2}[\Gamma-F_{0}] \equiv L_0
\end{equation}
where
$
F_{r}\equiv G_{r}+H_{r}
$

For $r\neq 0$, we have

\begin{equation}
\frac{\partial^{2}\bar f}{\partial \mathrm{Re}\ \chi_r\partial
\mathrm{Re}\ \chi_s}= A^{2}[\Gamma+ \mathrm{Re}\ F_{r}] \equiv
L_{r}(11) \delta_{rs} ,
\end{equation}

\begin{equation}
\frac{\partial^{2}\bar f}{\partial \mathrm{Im}\ \chi_r\partial
\mathrm{Im}\ \chi_s}= A^{2}[\Gamma- \mathrm{Re}\ F_{r}] \equiv
L_{r}(22) \delta_{rs} ,
\end{equation}

\begin{equation}
\frac{\partial^{2}\bar f}{\partial \mathrm{Re}\ \chi_r\partial
\mathrm{Im}\ \chi_s}= A^{2}[ \mathrm{Im}\ F_{r}] \equiv  L_{r}(12)
\delta_{rs} = L_{r}(21) \delta_{rs}
\end{equation}

The $\sigma q$, $\sigma\lambda$ and $\sigma\chi$ crossed terms vanish for $\sigma =0$:

\begin{equation}
\frac{\partial^{2}\bar f}{\partial\sigma^{\alpha}\partial
q^{\gamma\beta}_{r}}= \frac{\partial^{2}\bar f}{\partial\sigma^{\alpha}\partial
\lambda}=\frac{\partial^{2}\bar f}{\partial\sigma^{\alpha}\partial
\chi_{r}}= 0
\end{equation}

The only non-vanishing $q\lambda$ and $q\chi$ terms are the ones for
which $r = 0$. These are respectively,

\begin{equation}
\frac{\partial^{2}\bar f}{\partial \lambda \partial
q^{\alpha\beta}_{0}} = i\frac{\partial^{2}\bar f}{\partial
m^{2}\partial q^{\alpha\beta}_{0}}=-iAC_{0}
\end{equation}

\begin{equation}
\frac{\partial^{2}\bar f}{\partial\chi_{0}\partial
q^{\alpha\beta}_{0}}=-A^{2}C_{0}
\end{equation}
where
\begin{equation}
C_{0}\equiv\frac{H_{0}}{n(n-1)} \label{cr}
\end{equation}
The $r\neq 0$ terms vanish because $\bar q_r =0$ (and consequently
$H_r=0$) for $r\neq 0$.

Finally, the $\lambda\chi$ terms are
\begin{equation}
\frac{\partial^{2}\bar f}{\partial \lambda \partial \mathrm{Re}\
\chi_{r}}= i\frac{\partial^{2}\bar f}{\partial m^{2}\partial
\mathrm{Re}\ \chi_{r}}=-iA [ \mathrm{Re}\ F_{r}]
\end{equation}
and
\begin{equation}
\frac{\partial^{2}\bar f}{\partial \lambda \partial \mathrm{Im}\
\chi_{r}}= i\frac{\partial^{2}\bar f}{\partial m^{2}\partial
\mathrm{Im}\ \chi_{r}}=-iA [ \mathrm{Im}\ F_{r}].
\end{equation}

 The complete Hessian is as follows

\begin{widetext}
\begin{equation}
\mathbb{H}=\left(
  \begin{array}{ccccccccccccccccc}
    & &  & 0 & \cdots & &  &  &  &  &  &  &  &  & \cdots & 0  \\
    & \sigma_{\alpha\beta} &  & \vdots &  &  &  &  &  &  &  &  &  &  &  & \vdots  \\
    & &  & 0 & \cdots &  &  &  &  &  &  &  &  &  & \cdots & 0  \\
    0& \cdots& 0 &  &  &  &  &  &  &  &  & -iAC_{0} & -A^{2}C_{0} &  &  &    \\
    \vdots & & \vdots &  & K_{0} &  &  &  &  &  &  & \vdots & \vdots &  &  &    \\
    & &  &  &  &  &  &  &  &  &  & -iAC_{0} & -A^{2}C_{0} &  &  &    \\
    & &  &  &  &  & \ddots &  &  &  &  & \vdots &  & \ddots &  &    \\
    & &  &  &  &  &  &  &  &  &  & 0 &  &  & 0 &  0  & \\
    & &  &  &  &  &  &  & \widetilde{K}_{r} &  &  & \vdots &  &  & \vdots &    \\
    & &  &  &  &  &  &  &  &  &  & 0 &  &  & 0 &  0  &  \\
    & &  &  &  &  &  &  &  &  & \ddots & \vdots &  &  &  &  &  \ddots   \\
    & &  & -iAC_{0} & \cdots & -iAC_{0} & \cdots & 0 & \cdots & 0 & \cdots & \varphi & -iAF_{0} & \cdots & -iA\mathrm{Re}\ G_{r} &
     -iA \mathrm{Im}\ G_{r} \cdots   \\
    & &  &  &  &  &  &  &  &  &  &  &  &  &  &   \\
    & &  & -A^{2}C_{0} & \cdots & -A^{2}C_{0} &  &  &  &  &  & -iAF_{0} & L_0 &  &  &    \\
    & &  &  &  &  & \ddots &  &  &  &  & \vdots &  & \ddots &  &    \\
    \vdots& & \vdots &  &  &  &  & 0 & \cdots & 0 &  & -iA \mathrm{Re}\ G_{r} &  &  & {L}_{r}(11) & {L}_{r}(12)  \\
    \vdots& & \vdots &  &  &  &  & 0 & \cdots & 0 &  & -iA \mathrm{Im}\ G_{r} &  &  & {L}_{r}(21) & {L}_{r}(22) \\
    0& \cdots & 0 &  &  &  &  &  &  &  & \ddots & \vdots &  &  &  &  &  \ddots   \\

  \end{array}
\right)
\end{equation}
\end{widetext}

Here $\widetilde{K}_{r}$ is the matrix block with elements given by
(\ref{k1})-(\ref{k3}).

 A necessary and sufficient condition for the mean field solution to be
a local minimum is to have all the principal minors of the Hessian
positive. This would be equivalent to having all the eigenvalues of
the Hessian positive. The principal minors are the determinants of
the matrices obtained from the original matrix by successively
striping the last line and the last column, starting from the matrix
itself and ultimately reaching the (11) element. In the $n
\rightarrow 0$ limit we have the following set of principal minors
of $\mathbb{H}$: $D_\sigma$, $D_{q_0}$,..., $D_{q_r}$,...,
$D_{\lambda}$, $D_{\chi_0}$,..., $D_{\chi'_r}$, $D_{\chi''_r}$... .
In the previous expressions, the prime and double prime refer to the
two principal minor determinants generated by the sub-matrix
$L_r(ij)$.

We have carefully evaluated each of these determinants (see Appendix
C) in the limit $n \rightarrow 0$, for $\sigma =0$ (PM and SG
phases).

Defining

\begin{equation}
P_r \equiv \prod_{s = 1}^r\left[\frac{\Gamma^2 -
|G_s|^2}{\Gamma^2}\right] \label{pr}
\end{equation}
with
\begin{equation}
 P = \lim_{r\rightarrow\infty} P_r
\label{p}
\end{equation}
and
\begin{equation}
G \equiv \sum_{r}G_r = G_0 + 2 \sum_{r=1}^\infty \mathrm{Re}\ G_r
\label{g}
\end{equation}

(In Appendix D, we demonstrate that both $P$ and $G$ are finite and
positive).

we obtain
\begin{equation}
D_\sigma=1;
\end{equation}

\begin{equation}
D_{q_0}=\frac{1}{\Gamma}[(\Gamma-G_0)+H_0];
\label{q00}
\end{equation}

\begin{equation}
D_{q_r}=D_{q_0}P_r\ \ \ ;\ \ \  D_{\lambda}=D_{q_0}PG;
\label{qq}
\end{equation}

\begin{equation}
 D_{\chi_0}= A^2\left [ G_0^2 D_{q_0}+G \frac{(\Gamma-G_0)^2}{\Gamma}\right ]P;
 \label{dx0}
\end{equation}

$$
D'_{\chi_r}= A^2 [\Gamma + \mathrm{Re}\  G_r] D''_{\chi_{r-1}} +
$$
\begin{equation}
\frac{A^4}{\Gamma}(\Gamma-G_0)^2 [\mathrm{Re}\ G_r]^2 \left[
\prod_{s=1}^{r-1} A^2\left[\Gamma^2 - |G_s|^2\right ] \right ] P
 \label{dx1p}
\end{equation}

and

$$
D''_{\chi_r}=  \left[A^4 [\Gamma^2 - | G_r|^2]\right]
D''_{\chi_{r-1}} +
$$
\begin{equation}
\frac{A^2 }{\Gamma}(\Gamma-G_0)^2 A^2 |G_r|^2[\Gamma - \mathrm{Re}\
G_r]
 \left[
\prod_{s=1}^{r-1} A^2\left[\Gamma^2 - |G_s|^2\right ] \right ] P
 \label{dx2}
\end{equation}
 or equivalently,

$$
D''_{\chi_r}= \prod_{s=1}^{r} \left[A^4 [\Gamma^2 - | G_s|^2]\right]
D_{\chi_{0}} +
$$
\begin{equation}
\frac{A^2}{\Gamma}(\Gamma-G_0)^2 \sum_{s=1}^{r} A^4 |G_s|^2[\Gamma -
\mathrm{Re}\  G_s]
 \left[
\prod_{t\neq s}^{r} A^4\left[\Gamma^2 - |G_t|^2\right ] \right ] P
 \label{dx3}
\end{equation}

We see that all the principal minors, except $D_\sigma=1$ can be written in the form
\begin{equation}
\xi D_{q_0} + \eta [\Gamma-G_0]^2
\label{pm},
\end{equation}
where $\xi$ is a positive factor and $\eta$ is either positive or
zero. These properties follow from the fact that $\Gamma > |G_r|$,
for $r\neq 0$, as we demonstrate in Appendix B.

From (\ref{pm}), we see that $D_{q_0}$ is a key piece in the
evaluation of the principal minors. In the PM phase we have $\bar
q_0 =0,  H_0 = 0$ and $G_0 < \Gamma$, therefore it follows that
$D_{q_0} > 0$, hence all the principal minors are positive. In the
SG phase, conversely, $\bar q_0 > 0,  H_0 > 0$ and $G_0 = \Gamma$,
implying that again $D_{q_0} > 0$. It follows that all the principal
minors are positive also in the SG phase. At the transition, all
principal minors vanish, except for $D_\sigma=1$.

The above result establishes the thermodynamic stability of the SG
and PM phases obtained from our solution. Furthermore, we can see
the phase transition occurring by direct inspection of the principal
minors, since they vanish at the transition point. This result rules
out the usual instabilities found in the long-range interacting
replica-symmetric solutions. We can still have, of course,
meta-stability, which seems to be a characteristics of spin-glasses.
This would deserve further investigation. For this purpose, a
promising procedure would be to use the method of quantum annealing
\cite{das}, in order to find the absolute minimum of the free
energy.

\section{Concluding Remarks}

We have proposed a model for describing a short-range interacting, disordered quantum magnetic
 systems with SO(3) symmetry on a square lattice. The
random distribution of couplings is a Gaussian biased to an AF
coupling. A replica symmetric stable solution was obtained, which
clearly shows the existence of a stable genuine SG thermodynamical
phase, at a finite $T$. This can be seen directly from the solution,
but also by examining the principal minors of
 the Hessian matrix of the free energy, which are all positive, both in the PM and in the SG phases, but vanish at
the transition.

The use of $\Delta J \ll \bar J > 0$, allowed us to assume the
cancelation of the Berry phases. Relaxing this condition, we would
obtain a Chern-Simons term for the $A_\mu$-field in the final CP$^1$
version of the model. This system deserves a deeper investigation
but, presumably, the presence of this term will not modify the phase
structure found here.

Our solution takes into account the quantum fluctuations of the
CP$^1$ fields and therefore transcends the mean-field approximation.
This fact becomes evident, when we note that our solution does not
predict any ordered AF phase at $T\neq 0$, in agreement with the
Mermin-Wagner theorem, but contrary to what a mean field
approximation would yield.

The stable SG phase derives from a replica-symmetric solution. This
indicates that, in the case of short-ranged interactions, there is
no basic clash between the replica-symmetry of the SG solution and
its stability. We are naturally led to inquire, therefore, whether
the instability, which has been found in the SK solution, is
actually produced by the long-range interaction itself, rather than
by the replica symmetry it possesses.

The plots of the magnetic susceptibilities versus the temperature, exhibit the characteristic cusps,
experimentally found  at the
PM-SG transition, in materials exhibiting the SG phase.
By choosing a realistic value for the momentum cutoff $\Lambda$, we see that the cusps occur
precisely at the temperature values, which are observed experimentally. This provides clear evidence that
our model is really capable of describing realistic
SG systems and our results are not an artifact of the mean-field approximation.

Our model nicely describes the AF-SG transition, which occurs as we increase the amount of disorder, hence it will be
probably useful in the description of the corresponding transition in the high-Tc cuprates. We are currently investigating this point.


\begin{acknowledgments}

ECM would like to thank Curt Callan and the Physics Department of
Princeton University, where part of this work was done, for the kind
hospitality. This work was supported in part by CNPq and FAPERJ.
CMSC was supported by FAPERJ. We are grateful to P.R.Wells for the
help with the graphics.

\end{acknowledgments}

\vskip 10mm

\appendix
\renewcommand{\theequation}{\thesection.\arabic{equation}}
\setcounter{equation}{0}

\section{Determinant of the Quantum Fluctuations}

Let us evaluate here the determinant of the matrix of quantum
fluctuations $\mathbb{M}$ appearing in (\ref{seff1}). From
(\ref{mat}) and (\ref{acaoparadecompor11}) we obtain
\begin{eqnarray}
 \mathbb{M} =
\left(
\begin{array}{c}
K^{\alpha\beta} + A^{\alpha\beta}_{11} \ \ \ \ \   C^{\alpha\beta}_{12} \\ \\
C^{*\alpha\beta}_{21}\ \ \ \ \ \ \ \  K^{\alpha\beta} +
A^{\alpha\beta}_{22}
\end{array}\right)\;\label{mat1},
 \label{m}
 \end{eqnarray}
where, already in momentum-frequency space,
\begin{equation}
K^{\alpha\beta} =\left[|\vec k|^2 + \omega^2_r+m^2 \right
]\delta^{\alpha\beta},
\end{equation}
\begin{equation}
A^{\alpha\beta}_{ij} = A^{\alpha\beta} z^*_i z_j,
\end{equation}
and
\begin{equation}
C^{\alpha\beta}_{ij} = A^{\alpha\beta} z_i z_j.
\end{equation}
In these expressions,
\begin{equation}
A^{\alpha\beta} = \frac{2D}{\rho^2_s}\left [ [\bar q(\omega_r) -
\chi (\omega_r)] \delta^{\alpha\beta} - \bar q(\omega_r) \
\mathbb{C}^{\alpha\beta} \right ],
\end{equation}
where $\mathbb{C}^{\alpha\beta}$ is the $n \times n$ matrix with all
elements equal to one and $\chi (\omega_r)$ and $\bar q (\omega_r)$
are, respectively, the Fourier transforms of  $\chi(\tau-\tau')$ and
$\bar q(\tau-\tau')$.

We want to calculate
\begin{equation}
\ln \mathop{\mathrm{Det}} \mathbb{M}= \ln
\mathop{\mathrm{Det}}\limits^{}_{\vec
k\omega_r}\mathop{\mathrm{det}}\limits^{}_{\alpha\beta}
\mathop{\mathrm{det}}\limits^{}_{ij}\mathbb{M}, \label{seff11}
\end{equation}
where the three determinants run, respectively over the
momentum-frequency arguments of the fields, the replicas and the
$z$-field components. The first determinant can be easily evaluated
by diagonalizing (\ref{mat1}). We get
\begin{equation}
\ln \mathop{\mathrm{Det}} \mathbb{M}= \ln
\mathop{\mathrm{Det}}\limits^{}_{\vec
k\omega_r}\mathop{\mathrm{det}}\limits^{}_{\alpha\beta} \left [
K^{\alpha\beta} + \rho_s A^{\alpha\beta}   \right ]
 \label{seff111}
\end{equation}
 The two remaining determinants were evaluated in \cite{mm}. The one over the replicas yields exactly, for $n\rightarrow 0$,
\begin{equation}
\mathop{\mathrm{det}}\limits^{}_{\alpha\beta} \left [
K^{\alpha\beta} + \rho_s A^{\alpha\beta}   \right ] =
{\rm{N}}^n_{0}\left [1 - n\frac{\rm{N}_{1}}{\rm{N}_{0}}\right ]
\label{det1},
\end{equation}
where
$$\rm{N}_0=|\vec k|^2 + \omega^2_r+m^2 -\frac{2D}{\rho_s}\left(\chi_r-\bar q_r \right )$$
and
$$
\rm{N}_1=\bar q_r \left ( \frac{2D}{\rho_s} \right).
$$
Inserting (\ref{det1}) in (\ref{seff111}), finally, we can write the
last determinant as trace, which for $n\rightarrow 0$ reads
\begin{equation}
\ln \mathop{\mathrm{Det}} \mathbb{M}= n
\mathop{\mathrm{Tr}}\limits^{}_{\vec k\omega_r} \left [\ln
\rm{N}_{0} -\frac{\rm{N}_{1}}{\rm{N}_{0}}\right ] . \label{det11}
\end{equation}
This is given by
\begin{equation}
\ln \mathop{\mathrm{Det}} \mathbb{M}= n V
\sum_{\omega_{r}}\int\frac{d^{2}k}{(2\pi)^{2}}\left[\ln\left(k^{2}+M_r\right)
-\frac{A\bar q(\omega_{m})}{k^{2}+M_r }\right] , \label{det111}
\end{equation}
where $A = \frac{2D}{\rho_s}$ and $M_{r}$ is given by (\ref{mr}).

\section{ $\chi(\omega_r)$ and $ \bar q (\omega_r)$ for $r \neq 0$ }

\subsection{The Stationary Phase Equations}

 Let us define $6\pi \rho_s
[\chi(\omega_r) - \bar q (\omega_r)] \equiv \alpha +i \theta$, for
an arbitrary $r \neq 0$. Then, the mean field equation (\ref{m2})
yields the two equations
\begin{eqnarray}
2\theta \left [ \cosh \alpha - \cos \theta \right ] = \gamma \sin \theta
\label{mf1}.
\end{eqnarray}
and
\begin{equation}
m^2 + \omega_r^2 =  \frac{\Lambda^2\left[\cos \theta - e^{-\alpha}\right ]}{2\left [ \cosh \alpha - \cos \theta \right ]}
+\frac{\Lambda^2}{\gamma} \alpha
\label{mf2}.
\end{equation}

These equations imply $|G_r| < \Gamma$, as we demonstrate below.

\subsection{$|G_r| < \Gamma$}

Indeed, from (\ref{m2}), we have
\begin{equation}
\frac{\Lambda^2}{M_r} = e^{\alpha + i \theta}-1
\label{mf3}.
\end{equation}
This yields, from  (\ref{gr})
\begin{equation}
G_r = \frac{2\Gamma}{\gamma}\left [  \cosh(\alpha + i \theta) -1
\right ], \label{mf3}
\end{equation}
which implies
\begin{equation}
|G_r|^2 = \left (\frac{\Gamma}{\gamma}\right )^2\left [ 2 \left ( \cosh\alpha - \cos \theta \right ) \right ]^2.
\label{mf4}
\end{equation}
From (\ref{mf1}), we see that, for $\theta \neq 0$
\begin{equation}
|G_r| = \Gamma \left|  \frac{\sin \theta}{\theta} \right | < \Gamma
\label{mf5}
\end{equation}
For $\theta = 0$, conversely, $G_r$ is real and we see from Fig.1
that $\alpha_r < \alpha_0$ (for $r\neq 0$). Since $G_r$ is a
monotonically increasing function of $\alpha_r$, it follows that
$G_r < G_0$. Now, as we have seen, $G_0 \leq \Gamma$, hence $G_r <
\Gamma$ for $\theta = 0$. We conclude therefore that, for $r \neq
0$, we always have $|G_r| < \Gamma$.

For a non-vanishing $\bar q_r$, however, Eq. (\ref{IVa}) implies
$|G_r| = \Gamma$. We conclude, therefore, that (\ref{m2}), or
equivalently (\ref{mf1}) and (\ref{mf2}) will only admit solutions
for $r \neq 0$ when $\bar q_r = 0$. In this work, therefore we
always have $\bar q_{r\neq 0} = 0$. We also choose the $\theta =0$
solutions of (\ref{mf1}) and (\ref{mf2}) for all $r \neq 0$, which
imply real $\chi_r$'s.

\section{The Principal Minors of the Hessian}

\subsection{ The Determinant $D_\sigma$}

According to (\ref{ss}), the determinant of the ($\sigma\sigma$) $n \times n$ block of the Hessian, for finite $n$ is
given by
\begin{equation}
D_\sigma(n)= \left (\frac{M_0}{n} \right )^{n-1}\left[\frac{M_0}{n} - n A \frac{\bar q_0}{n} \right ]
\label{ds}.
\end{equation}
In the limit $n \rightarrow 0$, this gives
\begin{equation}
D_\sigma= \lim_{n \rightarrow 0} = 1 - n A \frac{\bar q_0}{M_0} = 1
\label{ds1},
\end{equation}
where we used the fact that
$$
\lim_{n \rightarrow 0} \left (\frac{M_0}{n} \right )^{n}=1.
$$

\subsection{ The Determinant $D_{q_0}$ }

From (\ref{qq})), we have the determinant of the ($q_0 q_0$) $n(n-1) \times n(n-1)$ block of the Hessian, for finite $n$, given by
\begin{equation}
D_{q_0}(n)= a_0^{[{n(n-1)-1}]}\left[a_0 - n(n-1) b_0 \right ]
\label{dq},
\end{equation}
where $a_0$ and $b_0$ are given by (\ref{ab}), for $r=0$.

In the $n \rightarrow 0$ limit, this gives
\begin{equation}
D_{q_0}= \lim_{n \rightarrow 0} \left ( \frac{n}{\Gamma}\right
)\left[\frac{\Gamma}{n} + \frac{G_0 - H_0}{ n(n-1)} \right ] =
\frac{1}{\Gamma}[(\Gamma-G_0)+H_0], \label{dq1}
\end{equation}
which is (\ref{q00}).

\subsection{ The Determinant $D_{q_r}, r\neq 1$}

 From
(\ref{k1})-(\ref{k3}), we can show that the determinant of the
$2n(n-1) \times 2n(n-1)$-dimensional, ($q_r q_r$) block of the
Hessian, for finite $n$, given by
$$
D^{(r)}_{q_r}(n)= \left[a_r^{[{2n(n-1)}]} + 2n(n-1) b_0
a_r^{[{2n(n-1)-1}]} |b_r|\right ] \times
$$
\begin{equation}
\left[a_r^{[{2n(n-1)}]} - 2n(n-1) b_0 a_r^{[{2n(n-1)-1}]}
|b_r|\right ]\label{dq1},
\end{equation}
where $a_r$ and $b_r$ are given by (\ref{ab}).

The limit $n \rightarrow 0$ can be taken in the same way as we did
in the previous subsection. Using the fact that $H_r =0$ for $r\neq
0$, we obtain,
\begin{equation}
D^{(r)}_{q_r}= \frac{\Gamma^2- |G_r|^2}{\Gamma^2}\label{dq2},
\end{equation}

Since the $q$-part of the Hessian is block-diagonal, we immediately
establish (\ref{qq}) for $D_{q_r}$. The limit $r \rightarrow \infty$
exists, as we show in Appendix C.

\subsection{ The Determinant $D_{\lambda}$ }

From (\ref{hessian}), we have that
\begin{equation}
D_{\lambda}(n)=  \varphi D_{q_\infty} +  A^2 n(n-1) C_0^2
a_0^{[{n(n-1)-1}]} \prod_{r\neq 0}^\infty \left [\frac{\Gamma^2 -
|G_r|^2}{\Gamma^2} \right ], \label{dl}
\end{equation}
where we used the fact that $c_r = 0$ for $r\neq 0$.

Now, using (\ref{ab}) and (\ref{cr}), we get
\begin{equation}
\lim_{n \rightarrow 0} A^2 n(n-1) C_0^2 a_0^{[{n(n-1)-1}]} = -\frac{H_0^2}{\Gamma}
\label{dl1}.
\end{equation}
Inserting in (\ref{dl}) and using (\ref{fi}), we immediately obtain
$D_\lambda$, considering that always $H_0 [\Gamma - G_0] =0 $.

\subsection{ The Determinant $D_{\chi_0}$ }

From (\ref{hessian}), we get
$$
D_{\chi_0}(n)= A^2 \left[\Gamma - (G_0 + H_0)\right ] D_{\lambda} +  A^2 F_0^2 D_{q_\infty}
$$
\begin{equation}
+ A^4 (2 F_0 - \varphi) n(n-1) C_0^2 a_0^{[{n(n-1)-1}]}P
\label{dx}
\end{equation}
Taking the limit $n \rightarrow 0$, using (\ref{dl1}) and (\ref{fi}), we obtain (\ref{dx0}), after a little algebra.

\subsection{ The Determinants $D'_{\chi_1}$ and $D''_{\chi_1}$}

From (\ref{hessian}), using the fact that  $H_r = 0$ for $r\neq 0$,
we get
$$
D'_{\chi_1}(n)= A^2 \left[\Gamma + \mathrm{Re} G_1\right ]
D_{\chi_0}
$$
$$
+  A^2(\mathrm{Re}\  G_1)^2\left[A^2 [\Gamma -G_0 -H_0]
D_{q_{\infty}} \right .
$$
\begin{equation}
\left . -A^4 2n(n-1) C_0^2 a_0^{[{2n(n-1)-1}]}\right ]P. \label{dx1}
\end{equation}
Using (\ref{dl1}), after some algebra we obtain, in the limit $n \rightarrow 0$,
\begin{equation}
D'_{\chi_1}= A^2 \left[\Gamma + \mathrm{Re}\  G_1\right ]
D_{\chi_0}+  A^4 (\mathrm{Re}\  G_1)^2\frac{[\Gamma -G_0
]^2}{\Gamma} P. \label{dx2}
\end{equation}

Following an analogous procedure, we obtain

$$
D''_{\chi_1}= A^4 \left[\Gamma^2 - | G_1|^2 \right ] D_{\chi_0}+
$$
\begin{equation}
A^4 | G_1|^2 \left[\Gamma - \mathrm{Re}\  G_1\right ]  \frac{[\Gamma
-G_0 ]^2}{\Gamma} P. \label{dx2}
\end{equation}

Considering the cases of $D_{\chi_2}$ and $D_{\chi_3}$, it is not
difficult to obtain, by induction, the general expression for
$D'_{\chi_r}$ and $D''_{\chi_r}$, eqs. (\ref{dx1p})-(\ref{dx3}).

\section{Finiteness of G and P}

\subsection{ Theorem: $0 < G < \infty$}

From (\ref{M}) and (\ref{gr}), or directly from (\ref{mf4}), after
choosing the solution $\theta =0$, we can write
\begin{equation}
G_r = \frac{T}{2\pi \Lambda^2} [\cosh(6 \pi \rho_s \chi_r) - 1] \geq
0 \label{gr1}
\end{equation}
As argued before, according to (\ref{m2}) or (\ref{M}), for large values of $r$, we have $\chi_r$ given by (\ref{xir}).
In this case,
\begin{equation}
G_r \approx \frac{T}{4\pi \Lambda^2}\left [\frac{\Lambda^2}{m^2 +
\omega_r^2}\right]^2 \label{gr2}
\end{equation}
and, for a sufficiently large but finite $N$,  $G$ can be written as
\begin{equation}
G = G_0 + 2 \sum_{r=1}^N G_r  + \frac{T}{2\pi \Lambda^2} \sum_{r>N}
\left [\frac{\Lambda^2}{m^2 + \omega_r^2}\right]^2 \label{gr3}
\end{equation}
The first two terms are obviously finite and positive. The third
term, is clearly smaller than
\begin{equation}
 \frac{T}{4\pi \Lambda^2} \sum_{r =0}^\infty \left [\frac{\Lambda^2}{m^2 + \omega_r^2}\right]^2,
\label{gr4}
\end{equation}
which is finite. It follows that $0 < G < \infty$.

\bigskip

\subsection{ Theorem: $0 < P < 1$}

From (\ref{pr}) and (\ref{p}), we have
\begin{equation}
 \ln P = \sum_{r =1}^\infty \ln\left[ 1 - \left (\frac{G_r}{\Gamma}\right )^2 \right ]
\label{gr5}
\end{equation}
Using the same idea of the previous subsection, we can write, for a
sufficiently large but finite $N$
\begin{equation}
 \ln P = \sum_{r =1}^N \ln\left[ 1 - \left (\frac{G_r}{\Gamma}\right )^2 \right ]
 -\frac{1}{\gamma^2}\sum_{r>N} \left [\frac{\Lambda^2}{m^2 +
 \omega_r^2}\right]^4
\label{gr6}
\end{equation}
The first term is obviously finite and negative. The modulus of the
second term is clearly smaller than
\begin{equation}
 \frac{1}{\gamma^2} \sum_{r =0}^\infty \left [\frac{\Lambda^2}{m^2 + \omega_r^2}\right]^4,
\label{gr7}
\end{equation}
which is finite. It follows that $0 < |\ln P| < \infty$, with $\ln P
< 0$. We conclude, therefore, that $ 0 < P < 1$.

\bibliography{apssamp}

\end{document}